%% file: paper.tex
\PassOptionsToPackage{dvipsnames,table}{xcolor}
\documentclass[acmsmall,screen]{acmart}

\usepackage[noend]{algorithmic}
\usepackage{algorithm}
\usepackage{graphicx}
\usepackage{tabularx}
\usepackage{textcomp}
\usepackage{array}
\usepackage{booktabs}
\usepackage{enumitem}
\usepackage{courier}
\usepackage{multirow}
\usepackage{colortbl}
\usepackage{subcaption}
\usepackage{tcolorbox}
\usepackage{listings}
\usepackage{makecell}
\usepackage{soul}
\usepackage{pifont}
\usepackage{svg}
\usepackage{url}
\usepackage{xspace}
\usepackage[toc,page]{appendix}
\usepackage{tikz}
\usetikzlibrary{shapes,arrows,decorations.pathreplacing,calc,shapes.misc,shapes.callouts,shapes.arrows,positioning,patterns}
\svgpath{{figures/}}

\DeclareTextFontCommand{\texttt}{\ttfamily\small}

\newcolumntype{L}[1]{>{\raggedright\arraybackslash}p{#1}}

\newcommand{\hlcolor}[2]{{\sethlcolor{#1}\hl{#2}}}

\definecolor{lsthcolorpink}{HTML}{FFB6C1}
\definecolor{lsthcolor}{HTML}{FFFFC5} 

\newcommand{\Grammarinator}{\textsc{Grammarinator}\xspace}
\newcommand{\ExplainFuzz}{\textsc{ExplainFuzz}\xspace}

\definecolor{mygreen}{RGB}{0,128,0}

\newcommand{\heatcolor}[1]{%
  \ifdim#1pt<0.00001pt\cellcolor{white!100}%
  \else\ifdim#1pt<0.05pt\cellcolor{orange!5}%
  \else\ifdim#1pt<0.1pt\cellcolor{orange!10}%
  \else\ifdim#1pt<0.2pt\cellcolor{orange!15}%
  \else\ifdim#1pt<0.3pt\cellcolor{orange!20}%
  \else\ifdim#1pt<0.4pt\cellcolor{orange!25}%
  \else\ifdim#1pt<0.5pt\cellcolor{orange!35}%
  \else\ifdim#1pt<0.6pt\cellcolor{orange!40}%
  \else\ifdim#1pt<0.8pt\cellcolor{orange!60}%
  \else\cellcolor{orange!80}%
  \fi\fi\fi\fi\fi\fi\fi\fi\fi#1}

\title{\ExplainFuzz: Explainable and Constraint-Conditioned Test Generation with Probabilistic Circuits}
\author{Annaëlle Baiget}
\email{annaelle.baiget@gmail.com}
\affiliation{\institution{UCLA}\city{Los Angeles}\state{CA}\country{USA}}

\author{Jaron Maene}
\authornote{This work was conducted while the author was a visiting student at UCLA.}
\email{jaron.maene@kuleuven.be}
\affiliation{\institution{KU Leuven}\city{Leuven}\country{Belgium}}

\author{Seongmin Lee}
\email{seongminlee@sigsoft.org}
\affiliation{\institution{UCLA}\city{Los Angeles}\state{CA}\country{USA}}

\author{Benjie Wang}
\email{benjiewang@g.ucla.edu}
\affiliation{\institution{UCLA}\city{Los Angeles}\state{CA}\country{USA}}

\author{Guy Van den Broeck}
\email{guyvdb@cs.ucla.edu}
\affiliation{\institution{UCLA}\city{Los Angeles}\state{CA}\country{USA}}

\author{Miryung Kim}
\email{miryung@cs.ucla.edu}
\affiliation{\institution{UCLA}\city{Los Angeles}\state{CA}\country{USA}}

\begin{document}

\begin{abstract}
  Understanding and explaining the structure of generated test inputs is essential for effective software testing and debugging. Existing approaches—including grammar-based fuzzers, probabilistic Context-Free Grammars (pCFGs), and Large Language Models (LLMs)—suffer from critical limitations. They frequently produce ill-formed inputs that fail to reflect realistic data distributions, struggle to capture context-sensitive probabilistic dependencies, and lack inherent explainability.
  We introduce \ExplainFuzz, a test generation framework that leverages Probabilistic Circuits (PCs) to learn and query structured distributions over grammar-based test inputs in an interpretable and controllable manner. Starting from a Context-Free Grammar (CFG), \ExplainFuzz compiles a grammar-aware PC and trains it on a corpus of existing inputs. New inputs are then generated via sampling. \ExplainFuzz utilizes the conditioning capability  of PCs to incorporate new test-specific constraints (e.g., a query must have \texttt{GROUP BY}). This enables constrained probabilistic sampling to generate high-quality inputs that satisfy the grammar and user-provided constraints.

  Our results show that \ExplainFuzz improves the coherence and realism of generated inputs, achieving a significant reduction in perplexity compared to probabilistic CFGs (pCFGs), {grammar-unaware} probabilistic circuits, and LLMs. By leveraging its native conditioning capability, \ExplainFuzz significantly enhances the diversity of inputs that satisfy a user-provided constraint. Compared to grammar-aware mutational fuzzing, \ExplainFuzz increases bug-triggering rates from 35\% to 63\% in SQL and from 10\% to 100\% in XML. These results demonstrate the power of {\em a learned input distribution} over mutational fuzzing, which is often limited to exploring the local neighborhood of seed inputs. Together, these capabilities highlight the potential of PCs to provide a foundation for grammar-aware, controllable test generation that can capture context-sensitive, probabilistic dependencies.
\end{abstract}

\maketitle

\input{chapters/1-introduction}

\input{chapters/2-related-work}
\input{chapters/3-approach-and-implementation}
\input{chapters/4-1-4-2experimental-design}

\input{chapters/4-rq2-conditioning}

\input{chapters/4-rq3-bugfinding}
\input{chapters/6-threat}
\input{chapters/7-conclusion}

\section{Data Availability}
\label{sec:data-avail}
We release all artifacts, including \ExplainFuzz source code, grammars,
custom concretizers, experimental data, and analysis scripts, at
\url{https://github.com/niMgnoeSeeL/ExplainFuzz/}.

\begin{acks}
This work is supported by the National Science Foundation under grant numbers 2426162, 2106838, and 2106404, and in part by funding from Amazon and Samsung. 
Jaron Maene received funding from the Flemish Government (AI Research Program), the European Research Council (ERC) under the European Union’s Horizon 2020 research and innovation program (Advanced Grant DeepLog No. 101142702), and was supported by a travel grant from the Research Foundation Flanders (FWO-Vlaanderen, V410925N). 
This work was funded in part by the DARPA ANSR, CODORD, and SAFRON programs under awards FA875023-2-0004, HR00112590089, and HR00112530141, NSF grant IIS1943641, and gifts from Adobe Research, Cisco Research, and Amazon. 
We thank the anonymous reviewers for their constructive feedback that helped improve this work.
\end{acks}

\newpage

\bibliography{ref}

\newpage

\appendix

\input{chapters/appendix.tex}

\end{document}

%% file: chapters/1-introduction.tex
\section{Introduction}

Generating high-quality test inputs is essential for effective software testing, particularly for compilers and databases that require complex, structured data. However, current methodologies suffer from a fundamental tradeoff between {\em generality} and {\em systematic control}. Traditional grammar-based fuzzers, such as \Grammarinator~\cite{10.1145/3278186.3278193}, can rapidly generate syntactically valid inputs but lack systematic control over the input distribution; because they rely on stochastic expansions or local seed mutations, they preserve only basic grammatical structures. Thus, they often produce unnatural or ill-formed inputs that fail to capture context-sensitive probabilistic dependencies. Conversely, specialized generators like SQLSmith~\cite{sqlsmith2018} produce high-quality, domain-specific inputs but do not generalize to other formats, as their input distributions are hard-coded into imperative logic and are not systematically tunable. Across these approaches, there is a pervasive lack of explainability and control regarding the input distributions. This leaves developers with no clear insight into which parts of the input space are being explored, nor a way to systematically adjust the generator's behavior using specific constraints during sampling.

Probabilistic context-free grammars (pCFGs) extend grammars with rule probabilities to bias generation toward common grammatically-correct structures. However, they lack the capacity for systematic control over {\em context-sensitive probabilistic dependencies}, as they assume independence between production rules. For example, in SQL, the likelihood of a \texttt{HAVING} clause is conditioned on the presence of a \texttt{GROUP BY} clause; a pCFG treats these as independent events, failing to capture the contextual patterns necessary for realistic input generation. Probabilistic models that are not aware of grammar structures, such as Probabilistic Circuits that represent Hidden Markov Models (PC-HMMs)\cite{ TractablePC2024,HMMtutorial}
or generic Probabilistic Circuits\cite{choi2020probabilistic} offer greater expressivity but make it impossible to guarantee syntactic validity, rendering them ineffective for testing compilers or database engines. Finally, while Large Language Models (LLMs) can often generate well-formed inputs, they act as uninterpretable {\em black boxes} that lack transparent control mechanisms beyond prompting.
Furthermore, they are computationally ill-suited for the high-volume, high-velocity, high-diversity sampling required in fuzzing campaigns.

We introduce \ExplainFuzz that leverages Probabilistic Circuits (PCs) to provide controllable and interpretable test generation that is \textbf{(1) grammar-aware}, can capture \textbf{(2) context-sensitive probabilistic dependencies}, and is \textbf{(3) systematically controllable via constraint conditioning}.

PCs are a class of generative models that represent complex probability distributions as computational graphs, offering a unique combination of expressivity and transparency~\cite{choi2020probabilistic,vergari2021compositional}. Unlike many deep-learning based generative models, PCs support probabilistic reasoning in linear time {with respect} to their size. This tractability makes them uniquely suited for fuzzing, as they allow for the rapid generation of inputs while providing fine-grained, systematic control over the distribution through native operations like {\em conditioning}\textemdash targeted sampling with a known constraint (e.g., generate a SQL query, but condition it so that it must contain a \texttt{JOIN} clause).

In \ExplainFuzz, we compile a Context-Free Grammar (CFG) to a grammar-aware probabilistic circuit, train the resulting circuit on seed inputs, and exploit its inference capabilities to perform two operations: \emph{probabilistic querying}\textemdash ``What is the probability that an input contains the token \texttt{JOIN}?'' and \emph{conditioned generation}\textemdash generate SQL queries that \emph{must} include \texttt{JOIN}. This combination of transparent querying and conditioned generation with a constraint enables both interpretability and control. This paper makes the following contributions:

\begin{itemize}[leftmargin=*]
  \item \textbf{Explainable and Controllable.} \ExplainFuzz constructs a grammar-aware probabilistic circuit (PC), trains it with seed inputs, performs probabilistic inference to produce explainable, grammar-structured, realistic inputs. Unlike traditional fuzzers, where the input distribution is a {\em black box} controlled only by a seed input set, PCs offer an interpretable view of the input space.

  \item \textbf{Improved Coherence and Realism.}
        \ExplainFuzz learns grammar-based input distributions that capture context-sensitive dependencies more effectively than probabilistic CFGs (pCFGs), probabilistic circuits without the knowledge of a grammar, or LLMs. Across seven domains, it reduces perplexity by a factor of 1.27$\times$ compared to pCFGs. This improvement demonstrates a superior capacity to capture implicit, context-sensitive probabilistic dependencies\textemdash dependencies that fall outside the formal scope of standard CFGs but are essential for generating realistic and coherent input structures.

  \item \textbf{Targeted Generation via PC Conditioning.}
        \ExplainFuzz generates targeted samples by conditioning the PC on user-specified constraints, systematically steering generation toward desired structures while preserving semantic coherence. Conditioning improves both bug coverage (+23.6\% for SQL, +17.5\% for XML) and input diversity, producing +21 (SQL) and +121 (XML) additional distinct bug-triggering inputs per bug on average (out of 10,000 samples) compared to unconditioned generation.
  \item \textbf{Bug-finding effectiveness.} Compared to \Grammarinator that performs grammar-aware mutational fuzzing, \ExplainFuzz's learned distribution improves average bug coverage from 35.3\% to 39.7\% in SQL and from 10\% to 82.5\% in XML, while producing +417 (SQL) and +1587 (XML) additional distinct bug-triggering inputs per seed set. This advantage arises because grammar-based mutation remains tightly bound to seed structures, whereas sampling from a learned probabilistic model explores a systematically broader region of the input space.

\end{itemize}

%% file: chapters/2-related-work.tex
\section{Background and Related Work}

\subsection{Grammar-Based Fuzzing}
Grammar-based fuzzing is a software testing technique that generates syntactically correct inputs based on a grammar. This method is particularly effective for testing parsers, interpreters, and compilers by producing complex, syntactically valid inputs.
Despite its strengths, grammar-based fuzzing struggles with context-aware probabilistic dependencies. For instance, in SQL, \Grammarinator~\cite{10.1145/3278186.3278193} might generate inputs like \texttt{SELECT DISTINCT '''' AS a FROM J4\$ AS O6, (* ) AS V4;} that are syntactically valid but semantically invalid\textemdash e.g., due to malformed sub-queries or illegal table aliases—rendering them unusable in practice. Beyond handling such explicit semantic violations such as illegal table aliases, test generation must also account for {\bf context-sensitive probabilistic dependencies}. For example, in SQL,
\texttt{GROUP BY} and \texttt{HAVING} are independent clauses, meaning you can have \texttt{GROUP BY} without \texttt{HAVING} and you can have \texttt{HAVING} without \texttt{GROUP BY}; however, the likelihood of \texttt{HAVING} is often increased when \texttt{GROUP BY} is used.

A common strategy to bias test generation is to use mutational fuzzing with a different seed set. However, such mutational fuzzing remains tied to local stochastic search around a seed set. It lacks an {\em explainable} or {\em controllable} means to bias the underlying input distribution because mutations merely explore local variations and rarely capture probabilistic dependencies within grammar-aware structures. This makes it difficult to diversify inputs, while retaining implicit dependencies present in real-world data.
In contrast, probabilistic modeling offers a principled way to learn a distribution over seed inputs. The model can capture context-sensitive probabilistic dependencies (e.g., how the presence of a \texttt{GROUP BY} influences the likelihood of a \texttt{HAVING} clause),  enabling \emph{controlled generation} that diversifies semantically coherent regions of the input space.

\subsection{Probabilistic Context Free Grammar Learning}

\begin{figure}[t]
  \centering
  \captionsetup{skip=2pt}
  \begin{minipage}[t]{0.67\textwidth}
    \begin{tcolorbox}[
        colback=gray!10,
        colframe=black,
        boxrule=0.4pt,
        arc=1mm,
        left=1mm,
        right=1mm,
        top=0.8mm,
        bottom=0.8mm,
        title=JSON Grammar Excerpt:,
        halign title=flush left,
        fonttitle=\bfseries\tiny
      ]
      \tiny
      \begin{verbatim}
value : obj | arr | STRING | NUMBER ;
obj   : '{' pair (',' pair)* '}' | '{' '}' ;
arr   : '[' value (',' value)* ']' | '[' ']' ;
pair  : STRING ':' value ;
\end{verbatim}
    \end{tcolorbox}
  \end{minipage}%
  \hfill
  \includegraphics[width=.3\textwidth,clip,trim=0 20 0 20]{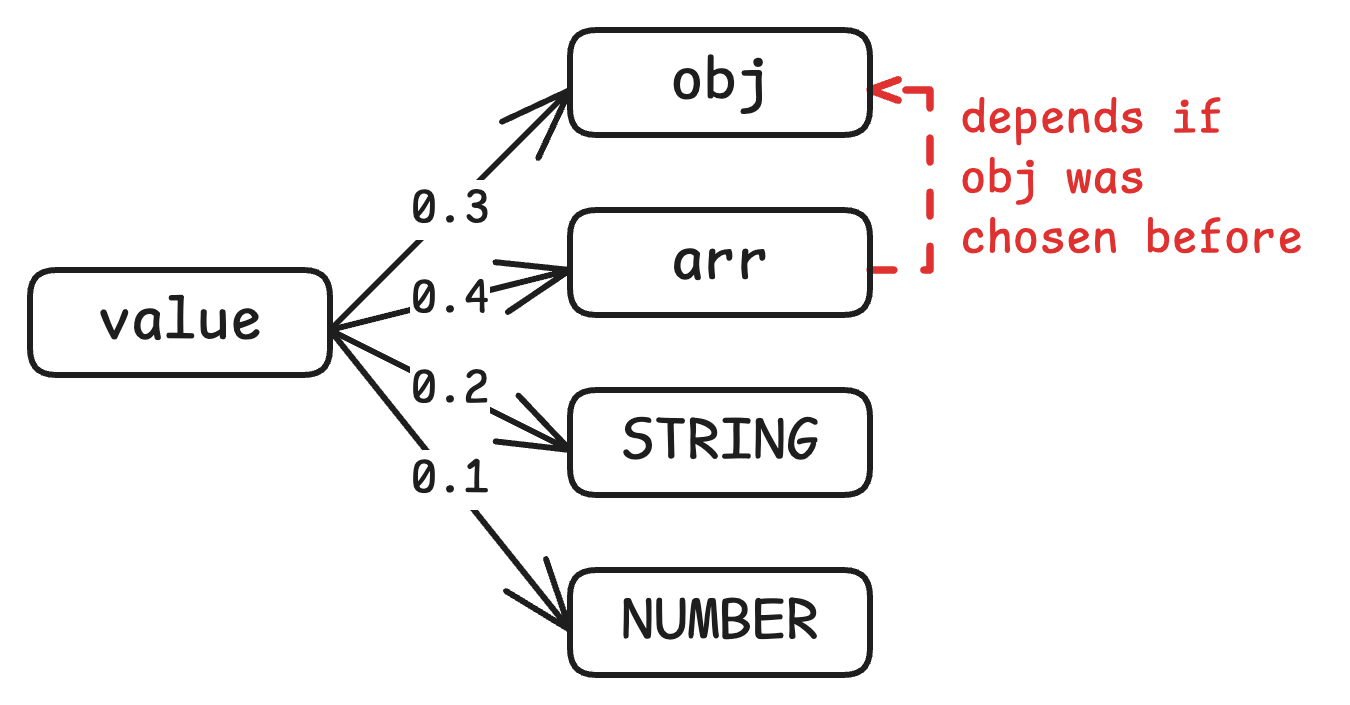}
  \noindent\rule{\textwidth}{0.4pt}
  \begin{minipage}[t]{0.67\textwidth}
    \begin{tcolorbox}[
        colback=gray!10,
        colframe=black,
        boxrule=0.4pt,
        arc=1mm,
        left=1mm,
        right=1mm,
        top=0.8mm,
        bottom=0.8mm,
        title=XML Grammar Excerpt:,
        halign title=flush left,
        fonttitle=\bfseries\tiny
      ]
      \tiny
      \begin{verbatim}
content          : element* ;
element          : nested_content | self_closing_tag ;
nested_content   : '<' Name attr* '>' content '<' '/' Name '>' ;
self_closing_tag : '<' Name attr* '/>' ;
\end{verbatim}
    \end{tcolorbox}
  \end{minipage}%
  \hfill
  \includegraphics[width=.3\linewidth,clip,trim=0 0 0 100]{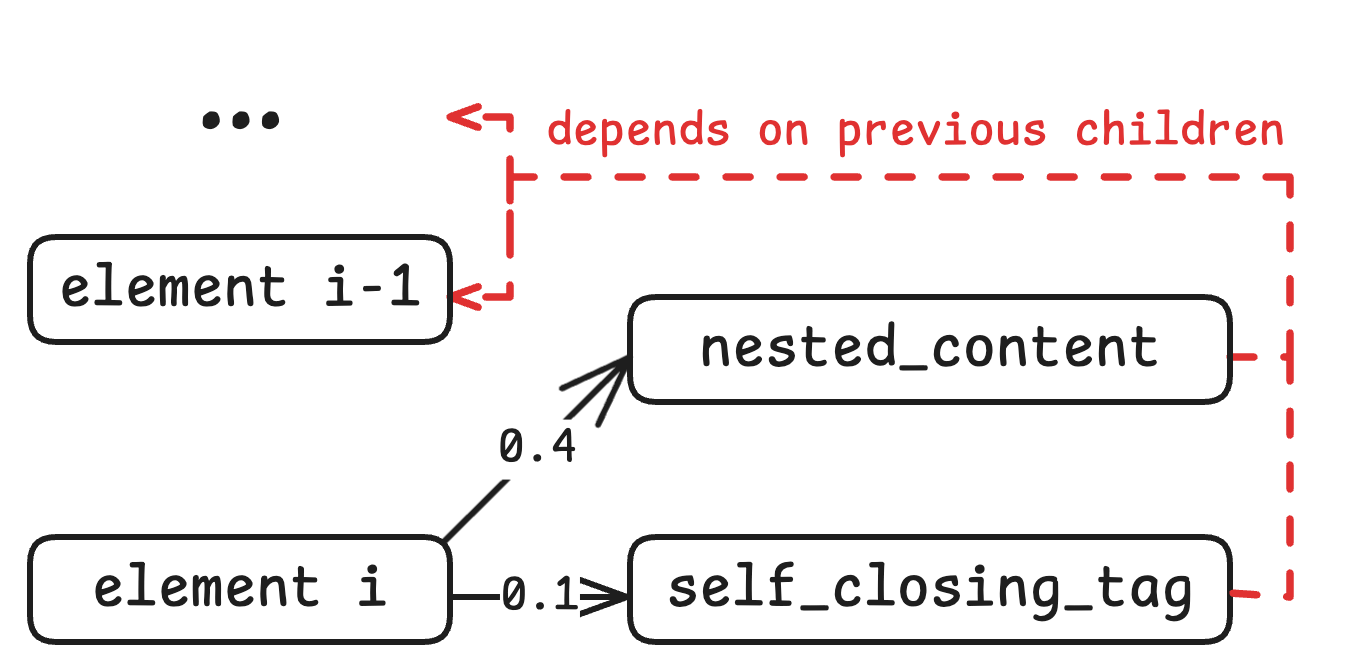}
  \caption{JSON and XML grammar excerpts (left) with context-sensitive generation graphs (right). Fixed weights illustrate pCFG probabilities; red arrows indicate context-sensitive probabilistic dependencies among inputs.
  }
  \label{fig:context_sensitive_examples}
\end{figure}

Probabilistic Context-Free Grammars (pCFGs) extend standard context-free grammars by associating each production rule with a probability, defining a distribution over possible derivations. These probabilities could be learned from a corpus of example inputs using the Expectation-Maximization (EM) algorithm, which iteratively optimizes the likelihood of the observed data under the model~\cite{Manning1999-mv,10.5555/1867406.1867499,9154602}. Probabilistic grammars have been used in fuzzing and program synthesis to bias input generation toward  likely structures.  For example, inputs-from-hell learns a probablistic context-free grammar from a corpus of inputs~\cite{9154602}.
However, standard pCFGs are fundamentally \textbf{limited by their independence assumption}: the probability of selecting a production rule is fixed and does not depend on previous choices or the broader derivation context such as ancestors or siblings. This limitation prevents pCFGs from capturing long-range, context-sensitive dependencies, which are crucial for realistic input modeling.

For example, in JSON (Figure~\ref{fig:context_sensitive_examples}, top), when expanding a \texttt{value} non-terminal, the choice between generating an \texttt{obj} or an \texttt{arr} cannot be conditioned on previously expanded siblings or the parent context under a pCFG; each expansion is treated independently. Similarly, in XML (Figure~\ref{fig:context_sensitive_examples}, bottom), deciding whether an \texttt{element} should be self-closing (\texttt{<tag/>}) or have nested content (\texttt{<tag>...\texttt{</tag>}}) may depend on the number or type of child \texttt{element}s already generated within the same parent, which a pCFG cannot account for.

Besides pCFGs, other probabilistic representations have been considered for test case generation. For example, LoadedDice \citep{TjoaOOPSLA25} utilizes probabilistic programs with learnable weights as a test case generator. However, the underlying representation based on binary decision diagrams is less efficient than CFGs in representing some languages.

\subsection{Probabilistic Circuits}

Probabilistic Circuits (PCs) are a family of tractable probabilistic models designed to represent and reason about complex data distributions in a structured and interpretable way~\cite{choi2020probabilistic,poon2011sumproduct,rahman2014cutset,vergari2021compositional,sidheekh2024building}.
Unlike black-box generative models such as neural networks or large language models (LLMs), PCs are built to support efficient probabilistic inference—allowing users to ask questions such as:
\emph{What is the probability of a specific input pattern?},
\emph{What features are most likely given partial information?}, or
\emph{How does changing a part of the input affect the overall distribution?}
These capabilities make PCs particularly useful for tasks that require explainable and controllable data generation, such as the synthesis of structured test inputs.  PCs have been successfully applied in domains such as computer vision~\cite{liu2022lossless,GargTPM25}, natural language processing~\cite{cheng2014language,zhang2024adaptablelogicalcontrollarge,yidouweng2025trace}, and knowledge graphs~\cite{loconte2023turn}. Recent work has explored how to make PCs more scalable for modeling complex, large-scale datasets \citep{loconte2024subtractive,wang2025inception,zhang2025monarch,seng2025scaling}.

A PC represents a probability distribution using a computational graph composed of three node types:
\emph{sum nodes} (capturing mixtures of alternatives),
\emph{product nodes} (capturing independent substructures),
and \emph{leaf nodes} (representing base distributions).
This structured representation provides a balance between expressiveness and tractability.
Under certain structural constraints—such as \emph{smoothness} and \emph{decomposability}—PCs enable exact computation of probabilistic queries (e.g., marginals, conditionals, and maximum a posteriori estimates) in time linear to the size of the circuit~\cite{choi2020probabilistic,wang2024algebraic,BroadrickICML25}.
This property is essential for applications where transparency and efficiency matter more than raw modeling capacity. 

For software testing, trained PCs offer two major advantages over traditional ML models:
(i) \textbf{Explainability}\textemdash their explicit structure allows testers to trace how different syntactic fragments or input tokens contribute to the overall likelihood of a test case; and
(ii) \textbf{Constraint-Conditioned Generation}\textemdash their probabilistic nature allows one to condition the PC on partial constraints, enabling targeted input generation (e.g., generating only inputs that satisfy having certain syntactic fragments). These features make PCs a compelling foundation for explainable and controllable test generation.

%% file: chapters/3-approach-and-implementation.tex
\section{\ExplainFuzz}

\subsection{Overview}

\begin{figure*}[htt]
  \centering
  \includegraphics[width=\textwidth]{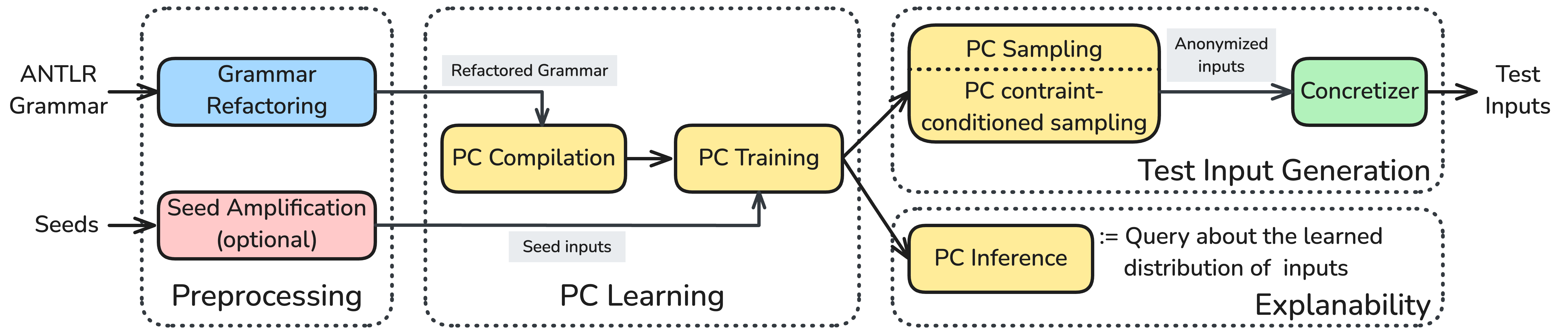}
  \caption{\ExplainFuzz: grammar-aware probabilistic circuit compilation, training, and sampling}
  \label{fig:overview}
\end{figure*}

This section describes the workflow of \ExplainFuzz, which consists of four stages. In the \textbf{Preprocessing} stage, \ExplainFuzz performs lightweight grammar refactoring.  In the \textbf{PC Learning} stage, it compiles a grammar to a grammar-aware probabilistic circuit and and trains the PC on the provided seed inputs. A user can use the resulting circuit in two ways: A user can inspect and understand the learned input distributions using the \textbf{PC Inference} capability. A user can then generate new inputs through PC sampling (generating a new data from the trained model), which are then concretized into executable forms.

\subsection{Preprocessing}

\paragraph{Grammar Refactoring}

Given an ANTLR grammar as input,
\ExplainFuzz applies lightweight grammar refactoring to ensure that ANTLR grammars can be compiled into a grammar-aware PC representation. It performs two key transformations:
\begin{itemize}[leftmargin=*]
  \item \textbf{Parsing and Splitting}: We use the ANTLR4 runtime to parse input grammars using generated parser and lexer components. Combined grammars are split into separate lexer and parser files to facilitate modular processing.
  \item \textbf{Literal Replacement}: Inlined literals in parser rules are replaced with named tokens, which are added to the lexer grammar. This makes the terminal alphabet explicit, enabling the PC to operate over symbolic tokens rather than raw literal strings and ensuring consistency across parsing, circuit construction, and concretization.
\end{itemize}

\paragraph{Seed Amplification and Anonymization}
\label{subsec:seed-input-generation}
To train the PC, we need training data (i.e., seed inputs). We optionally amplify this seed set by leveraging \Grammarinator.
Before training the PC, the generated inputs are \textit{anonymized}, by converting concrete lexemes to tokens.
The following shows an example anonymization:

\vspace{5pt}
\noindent
\resizebox{\linewidth}{!}{%
  \begin{tabular}{ccc}
    \textbf{Original}                                    &               & \textbf{Anonymized (Tokenized)} \\
    \midrule
    \texttt{SELECT name, age FROM users WHERE age > 42;} & $\rightarrow$ &
    \texttt{SELECT ID COMMA ID FROM ID WHERE ID GT Numeric SEMI}                                           \\
  \end{tabular}
}

\subsection{PC Compilation}

Our grammar-aware PC construction from a context-free grammar resembles bottom-up parsing algorithms such as the well-known CYK algorithm ~\cite{Zanzotto_2020}. Just as a bottom-up parser, we start with single terminal symbols and build up non-terminal derivations from there. Unlike a parser designed to recognize a string, the circuit construction models every possible input sequence up to a given length into account. This means we need to assume a maximum length of the input sequence.

\begin{algorithm}
  \caption{PC construction}\label{alg:cfg2pc}
  \footnotesize
  {
    \renewcommand{\algorithmicrequire}{\textbf{Input:}}
    \renewcommand{\algorithmicensure}{\textbf{Output:}}
    \begin{algorithmic}
      \REQUIRE Grammar $G$ in Chomsky normal form, maximum sequence length $n$, entry symbol $s_1$.
      \ENSURE PC $c$.
      \STATE Initialize empty circuit $c$.
      \FOR {symbol $s \in \textit{symbols}(G)$}
      \FOR {start $i \in [0..n-1]$}
      \FOR {end $j \in [i+1..n]$}
      \STATE Add a sum node $(s,i,j)$ to $c$.
      \ENDFOR
      \ENDFOR
      \ENDFOR
      \FOR {rule $(s_1 \leftarrow s_2\ s_3) \in \textit{rules}(G)$}
      \FOR {start $i \in [0..n-2]$}
      \FOR {middle $k \in [i..n-1]$}
      \FOR {end $j \in [k..n]$}
      \STATE Add a product node $(s_1, s_2, s_3, i,k,j)$ to $c$,
      \STATE with $(s_2, i, k)$ and $(s_3, k, j)$ as children.
      \ENDFOR
      \ENDFOR
      \ENDFOR
      \ENDFOR
      \FOR{product node $(s,i,j) \in c$}
      \STATE Add every sum node $(s,\_,\_,i,\_,j)$ as a child to $(s,i,j)$.
      \ENDFOR
      \STATE Add the root, a sum node with children $(s_1, 0, \_)$.
      \STATE Prune every node in $c$ that is not connected to the root.
    \end{algorithmic}
  }
\end{algorithm}

\begin{figure}
  \centering
  \begin{subfigure}[t]{0.42\textwidth}
    \raggedright
    \vspace{15pt}
    \subcaptionbox{ANTLR-style (EBNF) grammar}[0.95\linewidth][l]{%
      \small
      \(
      \begin{aligned}[t]
        \text{eq}    & : \text{digit } ( \text{PLUS digit} )^*; \\
        \text{digit} & :\text{ONE} \mid \text{TWO};             \\
        \text{ONE}   & : \texttt{"1"};                          \\
        \text{TWO}   & : \texttt{"2"};                          \\
        \text{PLUS}  & : \texttt{"+"};
      \end{aligned}
      \)
    }
    \par\vspace{15pt}
    \subcaptionbox{BNF-style grammar}[0.95\linewidth][l]{%
      \small
      \(
      \begin{aligned}[t]
        \text{eq}    \rightarrow~ & \text{ONE block} \mid \text{TWO block} \mid \text{eq block} \\
        \text{block} \rightarrow~ & \text{PLUS ONE} \mid \text{PLUS TWO}                        \\
        \text{ONE}   \rightarrow~ & \texttt{"1"}                                                \\
        \text{TWO}   \rightarrow~ & \texttt{"2"}                                                \\
        \text{PLUS}  \rightarrow~ & \texttt{"+"}
      \end{aligned}
      \)
    }
  \end{subfigure}
  \begin{subfigure}[t]{0.54\textwidth}
    \centering
    \vspace{0pt}
    \includegraphics[width=\linewidth]{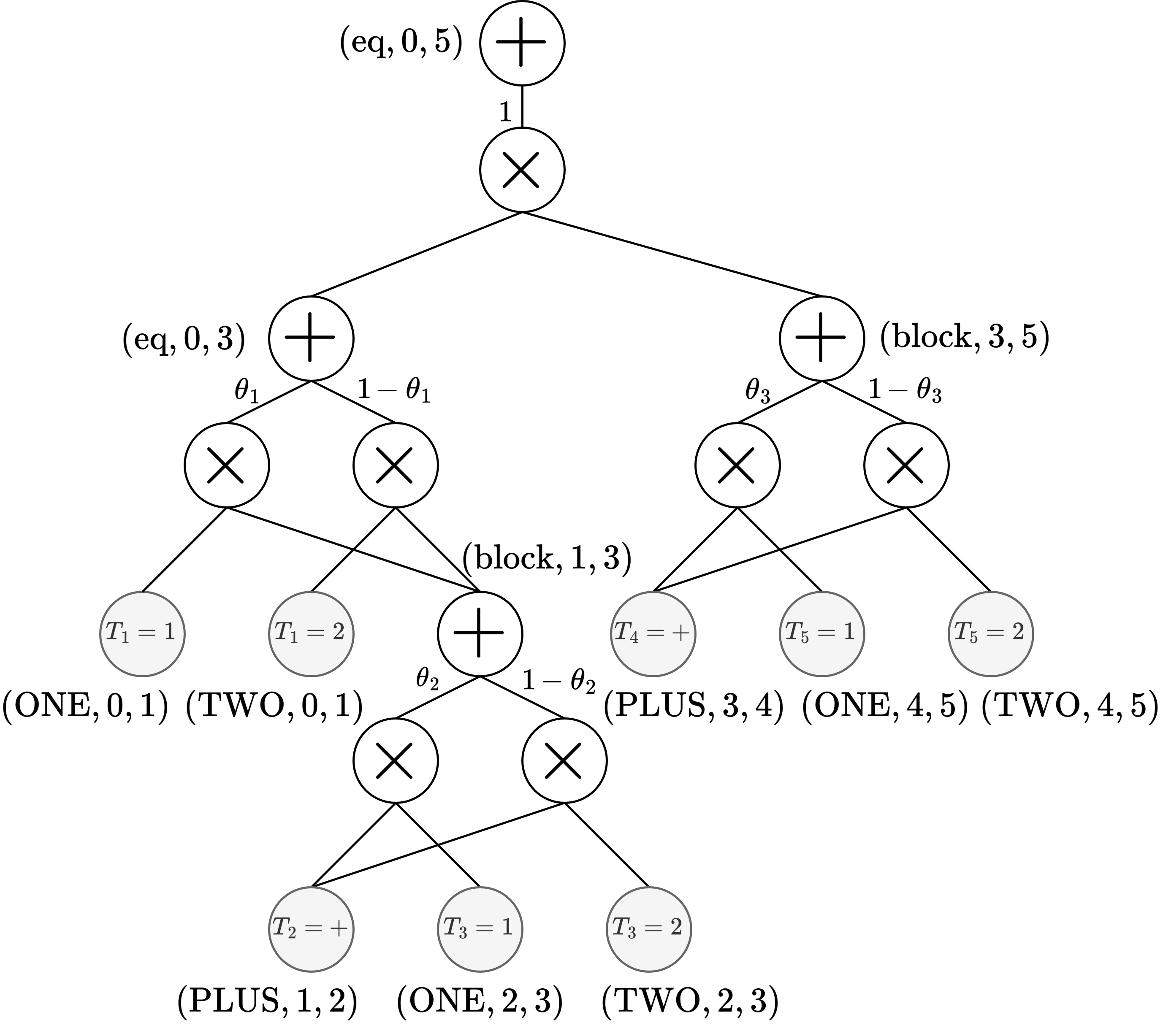}
    \caption{PC}
  \end{subfigure}
  \caption{ANTLR-style EBNF grammar of simple arithmetic expressions, a refactored BNF-style grammar, and the corresponding PC (sequence length 5) built from the BNF grammar. Observe that $\theta_2$ and $\theta_3$ would be the same value in a PCFG, while they may differ in the PC. }
  \label{fig:grammar-and-pc}
\end{figure}

Algorithm~\ref{alg:cfg2pc} provides pseudo-code for our grammar-aware PC construction. The circuit construction first adds a \texttt{sum} node for each non-terminal symbol in the grammar and each subset of the sequence (i.e., a starting and ending position). This sum node computes the probability that a subset of the string parses to a specific non-terminal. As children, this sum node has a \texttt{product} node for each rule that can produce the non-terminal at this sequence subset. These product nodes are in turn connected to all sum nodes that are required in the rule production.
Figure~\ref{fig:grammar-and-pc} illustrates the process of compiling a simple arithmetic ANTLR grammar to the resulting PC using Algorithm~\ref{alg:cfg2pc} with a maximum sequence length of 5.

From the training data set, we filter out all the inputs that have a number of tokens higher than the \textit{max\_length} parameter of the PC. This parameter is chosen for each domain to reflect the expected length of an input. Then, the PC is trained with the data to learn distributions over symbolic grammar tokens.

\subsection{Explainability with PC Inference}
\label{sec:inference}

A key benefit of PCs is their support for tractable inference, which enables users to not only generate samples but also to inspect, diagnose, and understand the underlying structure the model has learned. Before generating concrete tests, a user can use PC inference capability to understand learned distributions, to ask which tokens are over or under-represented, etc.

\begin{table}[t]
  \centering
  \caption{Examples of the four query classes supported by \ExplainFuzz.}
  \label{tab:pc-inference-queries}
  \resizebox{\linewidth}{!}{
    \begin{tabular}{lll}
      \toprule
      \textbf{Query type} & \textbf{Form}                                                                   & \textbf{Example / interpretation} \\
      \midrule
      EVI  (Evidence)     & Probability of a specific input.                                                &
      \makecell[l]{
      Probability assigned to this exact anonymized input:                                                                                      \\
        $\qquad P(\text{``SELECT ID FROM ID ORDER BY ID ;''})$
      }                                                                                                                                         \\
      \midrule
      MAR  (Marginal)     & (i) Token marginal: probability of token $t$ appearing in an input.             &
      Probability that the input contains at least one \texttt{JOIN}: $P(\text{``JOIN''})$                                                      \\
                          & (ii) Span marginal: probability of a contiguous sequence at position $p$.       &
      \makecell[l]{
      Probability that \texttt{GROUP BY ID} starts at position~6:                                                                               \\
        $\qquad P([\text{``GROUP'', ``BY'', ``ID''}],\, \text{position}=6)$
      }                                                                                                                                         \\
      \midrule
      COND (Conditional)  & (i) Token conditional: probability of token $t_2$ given token $t_1$.            &
      Probability of \texttt{HAVING} given \texttt{GROUP}: $P(\text{``HAVING''} \mid \text{``GROUP''})$                                         \\
                          & (ii) Next-token conditional at position $p$.                                    &
      \makecell[l]{
      Probability that \texttt{BY} follows \texttt{GROUP} at position~8:                                                                        \\
        $\qquad P([\text{``GROUP'', ``BY''}] \mid (\text{``GROUP''},\, \text{position}=8))$
      }                                                                                                                                         \\
                          & (iii) Following-token conditional after a given span at position $p$.           &
      \makecell[l]{
      Probability that \texttt{GROUP} appears after \texttt{JOIN ID ON} at position~12:                                                         \\
        $\qquad P(\text{``GROUP''} \mid ([\text{``JOIN'', ``ID'', ``ON''}],\, \text{position}=12))$
      }                                                                                                                                         \\
      \midrule
      MMAP (MAR+MAP)      & (i) Most likely token appearing anywhere after token $t$.                       &
      Most likely token to appear after \texttt{HAVING}.                                                                                        \\
                          & (ii) Most likely token appearing immediately after token $t_1$ at position $p$. &
      Most likely token immediately after \texttt{JOIN} when \texttt{JOIN} is at position~8.                                                    \\
      \bottomrule
    \end{tabular}}
\end{table}

\ExplainFuzz addresses these problems through four classes of tractable inference queries\textemdash complete evidence, marginal, conditional, and marginal MAP\textemdash each mapped directly to a user-facing analysis task. Table~\ref{tab:pc-inference-queries} illustrates each category with concrete SQL examples. Evidence (EVI) query can answer, ``How likely is this specific scenario?'' Marginal (MAR) query answers, ``What is the probability of X, ignoring Y?'', by calculating the probability of a subset of variables while marginalizing out the others. A maximum a posteriori (MAP) query answers ``What is the most likely assignment to these variables?'' Conditional query (COND) can answer, ``What is the probability of some variables given that we already know the values of others?'' These PC inference queries enable \ExplainFuzz to serve as an \textbf{explanatory tool} beyond a black-box fuzzer (i.e., sampler). A user can examine a learned distribution, understand generated inputs, and refine the input-generation process, enabling principled adjustments before downstream test generation (i.e., PC sampling).

\subsection{Test Input Generation with PC}
\paragraph{PC Sampling}
Beyond PC inference queries, the trained PC can be used for generating inputs via sampling from the learned distribution. This allows us to explore the space of grammar-conforming inputs that reflect the context-sensitive probabilistic dependencies within grammatical structures.
Moreover, the PC supports \textbf{constraint-conditioned sampling}, enabling generation under specific constraints. For instance, one can sample inputs that include a given token either anywhere in the sequence or at a specific position. It is also possible to enforce the presence of fixed subsequences of tokens, while allowing the rest of the input to vary. This flexible constraint-based conditioning capability makes PCs a powerful tool for targeted generation with systematic control, as opposed to providing a different seed set, and hoping that mutational fuzzing will retain context-sensitive probabilistic dependencies.

\paragraph{Concretization of Tokens}

\begin{table}[t]
  \centering
  \caption{Example concretization of anonymized tokens to concrete string.}
  \label{tab:concretization_examples}
  \scriptsize
  \resizebox{\linewidth}{!}{%
    \begin{tabular}{l l l}
      \toprule
      \textbf{Domain}      & \textbf{Form} & \textbf{Conversion}                                                                   \\
      \midrule
      \multirow{2}{*}{SQL} & PC sample     &
      \texttt{\scriptsize SELECT ID COMMA ID FROM ID WHERE ID EQUAL Integral ORDER BY ID ASC SEMI EOF}                             \\
                           & Concretized   &
      \texttt{\scriptsize SELECT salary, performance\_score FROM employees WHERE salary = 80000 ORDER BY performance\_score ASC ;} \\ \midrule
      \multirow{2}{*}{XML} & PC sample     &
      \texttt{\scriptsize OPEN Name Name EQUALS STRING Name EQUALS STRING SLASH\_CLOSE EOF}                                        \\
                           & Concretized   &
      \texttt{\scriptsize <file path="auth\_service\_xml/data/resources.api.txt" name="api"/>}                                     \\
      \bottomrule
    \end{tabular}
  }
\end{table}

The PC operates over tokens (i.e., anonymized symbols), so its generated sample must be replaced concrete values (e.g., strings, numbers, identifiers, XML elements). To produce executable and human-readable inputs, we apply a lightweight concretization method that maps placeholder tokens to valid literals. In our evaluation, we provide a custom concretizer implementation for each domain: SQL and XML; however, replacing tokens to concrete values can be easily done with LLMs as well. In brief, the SQL concretizer grounds identifiers in the target schema and fills literals with type-consistent values (e.g., numeric ranges, string formats), while the XML concretizer instantiates tags and attributes using schema rules and project metadata (e.g., entity names, file paths) to keep outputs valid.\footnote{Engineering details and artifacts are available in the open-source release (Section~\ref{sec:data-avail}).} Table~\ref{tab:concretization_examples} shows an example of replacing anonymized tokens to concrete strings in the SQL and XML domains.

%% file: chapters/4-1-4-2experimental-design.tex
\section{Evaluation}
\label{sec:evaluation}

We structure our evaluation with the following three research questions:

\begin{itemize}[leftmargin=*,itemsep=0.5em]
  \item \textbf{RQ1 Coherence and Realism: How well does \ExplainFuzz generate coherent and realistic test inputs compared to baseline models?}
        A model that accurately captures the distribution of real inputs\textemdash including context-sensitive probabilistic dependencies\textemdash will generate more coherent and realistic test inputs. We measure this using perplexity on a held-out test set across seven domains, comparing \ExplainFuzz against a probabilistic context-free grammar (pCFG), a grammar-unaware probabilistic circuit (PC-HMM), and a large language model (LLM).

  \item \textbf{RQ2 Impact of Targeted Sampling: To what extent does probabilistic conditioning improve \ExplainFuzz's ability to target specific structures while maintaining input diversity?}
        Many bugs are triggered only by rare syntactic or semantic structures; conditioning aims to steer generation toward these edge cases without sacrificing structural validity or global diversity. We evaluate the impact of conditioning by comparing the diversity of bug-triggering inputs generated by conditioned \ExplainFuzz against its unconditioned counterpart.
  \item \textbf{RQ3 Bug Triggering Input Diversity: Does \ExplainFuzz's learned distribution improve bug-triggering rates over \Grammarinator?}
        We assess whether \ExplainFuzz’s probabilistic model produces higher-quality test inputs—defined as those that are more executable and effective at triggering bugs—relative to a grammar-based mutational fuzzer \Grammarinator. To ensure a fair comparison, we evaluate both systems across multiple seed sets of varying size and representativeness, isolating the influence of initial seeds on generator performance.
\end{itemize}

\subsection{RQ1 (Coherence and Realism)}

We train and compare four models: (1) grammar-aware PC, (2) grammar-unaware PC, (3) LLM, and (4) pCFG. These models are all trained on the same seed-based dataset for each domain (Section~\ref{subsec:seed-input-generation}).

\begin{itemize}[leftmargin=*]
  \item The first baseline is a \textbf{probabilistic context-free grammar (pCFG)}, whose rule probabilities are estimated from the same corpus. As the canonical grammar-based statistical model, pCFGs serve to test whether \ExplainFuzz's grammar-aware PC can capture context-sensitive probabilistic dependencies. Since PCs can enforce global dependencies and share parameters across substructures, we hypothesize that \ExplainFuzz will achieve lower perplexity than pCFGs, particularly in domains with nested or cross-cutting dependencies.
  \item The second baseline is an \textbf{\emph{grammar-unaware} probabilistic circuit (PC-HMM)}: a probabilistic circuit that represents a {Hidden Markov Model (HMM)} and thus does not consider syntactic constraints; This baseline mirrors the language modeling component of constrained-decoding, Ctrl-G~\cite{zhang2024adaptablelogicalcontrollarge}, but without grammar awareness. We include it to assess the importance of grammatical structure: while PC-HMM captures local sequential patterns, we expect it to underperform \ExplainFuzz on hierarchical, structural dependencies, though still outperforming pCFGs due to its higher representational capacity.
  \item The final baseline is a pre-trained \textbf{LLM} ({GPT-2} model) prompted with five anonymized seed queries. This comparison evaluates (1) whether a general-purpose LLM can approximate the distribution of anonymized, domain-specific inputs, and (2) the limitations of LLMs in narrow structured domains. We expect GPT-2 to struggle, as anonymization removes concrete string cues and induces distributions not represented in its pre-training data, hindering accurate likelihood estimation between components within grammatical structures.
\end{itemize}

We implement the circuit construction in Python using the KLay library~\cite{maene2025klay}, enabling GPU-accelerated training. As a sanity check, we compare PC-estimated probabilities against empirical probabilities computed from the training data used to fit the circuit. The estimates showed low error and closely matched the data distribution, indicating the PC was trained correctly to capture the underlying distribution. All experiments are run on a MacBook Air (M3) with 16 GB RAM and 8-core processor.

Specifically for preparing a training set and test data set, we use \Grammarinator in mutational mode with the provided seed inputs and apply grammar-based mutations to amplify the initial seeds, resulting in 10,000 inputs. We split this dataset into a training set of 9,000 inputs and a held-out test set of 1,000 inputs. We then measure the \textbf{average perplexity per token}, a standard measure of a probabilistic model's predictive uncertainty:
\[
  \text{Perplexity} = \exp\left(- \frac{1}{T} \sum_{i=1}^{N} \log P(x_i) \right)
\]
where $x_i$ is a test query, $P(x_i)$ is the probability assigned by the model, and $T$ is the total number of tokens across all test queries. Lower perplexity indicates that a model assigns higher likelihood to observed data, hence better capturing the true input distribution. We evaluate all models across seven domains: {SQL}, {JANUS}, {REDIS}, {B}, {CSV}, {HTML}, and {JSON}. For each domain, we set a maximum input length (\texttt{max\_length}) to control the complexity of generated inputs, as detailed in the table in Figure~\ref{fig:perplexity_overview}.

\begin{figure*}[t]
  \centering
  \begin{subfigure}[t]{0.53\textwidth}
    \centering
    \vspace{0pt}
    \resizebox{\linewidth}{!}{%
      \begin{tabular}{c|c|cccc}
        \toprule
        \multirow{2}{*}{\textbf{Domain}} & \textbf{Max}    & \multirow{2}{*}{\textbf{LLM}} & \multirow{2}{*}{\textbf{pCFG}} & \multirow{2}{*}{\textbf{PC-HMM}} & \textbf{Grammar-aware PC} \\
                                         & \textbf{length} &                               &                                &                                  & \textbf{(\ExplainFuzz)}   \\
        \midrule
        B                                & 40              & 7.19                          & 2.43                           & 1.53                             & \textbf{1.51}             \\
        CSV                              & 45              & 2.83                          & 1.46                           & 1.20                             & \textbf{1.15}             \\
        HTML                             & 50              & 9.03                          & 1.31                           & 1.26                             & \textbf{1.20}             \\
        JANUS                            & 30              & 15.05                         & 2.53                           & \textbf{1.56}                    & 1.61                      \\
        JSON                             & 50              & 6.37                          & 1.88                           & 1.37                             & \textbf{1.32}             \\
        REDIS                            & 15              & 2744.53                       & 2.13                           & \textbf{2.09}                    & 2.19                      \\
        SQL                              & 35              & 280.93                        & 1.52                           & 1.41                             & \textbf{1.34}             \\
        \bottomrule
      \end{tabular}}
    \label{tab:perplexity_domain_comparison}
  \end{subfigure}
  \hfill
  \caption{Perplexity comparison between grammar-aware PC, probabilistic CFG, grammar-unaware PC (PC-HMM), and LLM across domains.}
  \label{fig:perplexity_overview}
\end{figure*}

Figure~\ref{fig:perplexity_overview} shows the average perplexity achieved by each model across seven domains: {SQL}, {JANUS}, {REDIS}, {B}, {CSV}, {HTML}, and {JSON}. Lower perplexity indicates that a model better captures the true input distribution—including context-sensitive dependencies—and thus generates more coherent and realistic inputs. We observe that the grammar-aware PC model (\ExplainFuzz) achieves the lowest perplexity in five out of seven domains (SQL, B, CSV, HTML, and JSON). In contrast, the grammar-unaware PC model (PC-HMM) performs best in JANUS and REDIS, where \ExplainFuzz ranks second and last, respectively.

The PC-HMM's advantage on JANUS and REDIS likely stems from their data characteristics. REDIS has low syntactic regularity and high lexical variability: commands are short keyword patterns (e.g., \texttt{SET key value}, \texttt{INCR counter}, or \texttt{LPUSH list item}) rather than deep hierarchies, reducing the benefits of grammar-aware modeling. While JANUS has a rich grammar, the max-length constraint yields limited syntactic diversity, making local token transitions highly predictive; the flat HMM can thus fit these patterns without being constrained by infrequent or unused grammar rule productions. Although PC-HMM can occasionally win on perplexity, it lacks grammatical constraints and may thus generate syntactically invalid inputs. We evaluated 100,000 samples generated from a PC-HMM in the MLIR domain and observed a parsing rate of only 12.9\%. This result \textbf{underscores the need for grammar-aware PCs},  to produce syntactically valid inputs capable of bypassing initial parsing checks.

Compared to pCFGs, \ExplainFuzz benefits from probabilistic circuits that can duplicate and specialize rules, capturing context-sensitive dependencies beyond fixed production probabilities and yielding lower perplexity in most domains. Despite its vast pretraining, the LLM performs worse across all domains. Its general-purpose training objective is mismatched to tokenized inputs in a specialized domain, which further inflates perplexity.

\paragraph{\textbf{RQ1 Result Summary. }}
\ExplainFuzz generates more coherent and realistic test inputs than the baselines by capturing context-sensitive dependencies that pCFGs, grammar-unaware PCs, and LLMs fail to capture adequately. It achieves the lowest perplexity in five of seven domains, demonstrating a superior capacity to learn the probabilistic dependencies within grammatical structures required to model complex, real-world hierarchical inputs.

\subsection{Experimental Setup for RQ2 and RQ3}
Both RQ2 and RQ3 use the shared experimental setup—including domains, testbeds, bug oracles, and seed sets.

\paragraph{Shared Testbed Infrastructure}

We evaluate \ExplainFuzz on two representative domains: SQL and XML. These domains were chosen primarily because we developed custom concretizers for them, which allow us to convert tokenized  samples from PC into concrete, executable inputs. They are also well-suited for evaluating grammar-aware PC model: both domains have rich, hierarchical structure and conditional dependencies, allowing us to demonstrate the \ExplainFuzz's ability to execute inference queries and to generate new samples conditioned on specific constraints.

For each domain, we develop a testbed featuring a system under test (SUT) and synthetic bug oracles modeled after real-world bug scenarios. These bug oracles identify inputs that mimic known vulnerabilities in SQL and XML systems, allowing us to evaluate how effectively a fuzzer triggers various bugs. In addition, we design \textbf{multiple seed sets}. This is to compare grammar-based mutational fuzzing with seeds with sampling from grammar-aware PC. While both strategies can be used to bias grammar-aware input generation with seeds, grammar-aware PC can uniquely capture context-sensitive probabilistic dependencies. To evaluate the ability of \Grammarinator and \ExplainFuzz to discover previously unseen unseen bugs, we ensure none of the seed sets include inputs that can trigger bugs on their own.  Below, we summarize the design of each testbed; full details\textemdash including schemas and project layouts\textemdash are provided in our open-source artifacts (Section~\ref{sec:data-avail}).

\vspace{.5em}
\subparagraph{\textbf{SQL testbed}.}

\begin{table*}[t]
  \caption{(left) Characteristics of the four seed sets used in our evaluation for the SQL domain. (right) Marginal probability of key SQL constructs across four subdomains; the last row, `nested select`, is computed as $P(\text{"SELECT"}) \cdot P(\text{"SELECT"} \mid \text{"SELECT"})$.}
  \label{tab:seed_sets}
  \centering
  \begin{subtable}[t]{0.64\textwidth}
    \centering
    \resizebox{\linewidth}{!}{%
      \begin{tabular}{l >{\centering\arraybackslash}p{3 cm} > {\centering\arraybackslash}p{1cm} >{\centering\arraybackslash}p{4cm} >{\centering\arraybackslash}p{4cm}}
        \toprule
        \textbf{ID} & \textbf{Seed Set}       & \textbf{Size} & \textbf{Main Property}              & \textbf{Expected Effect}         \\
        \midrule
        SQL1        & Homogeneous Structure   & 50            & Repeated query templates            & Low structural diversity         \\
        \midrule
        SQL2        & Homogeneous Column      & 50            & Frequent reuse of sensitive columns & Column bias (for \Grammarinator) \\
        \midrule
        SQL3        & Very Small              & 10            & Minimal coverage                    & Underfitting, low coverage       \\
        \midrule
        SQL4        & Diverse (High-Coverage) & 50            & Broad structural variety            & Best-case diversity              \\
        \bottomrule
      \end{tabular}}
  \end{subtable}
  \hfill
  \begin{subtable}[t]{0.35\textwidth}
    \centering
    \resizebox*{\linewidth}{!}{%
      \begin{tabular}{lcccc}
        \textbf{Literal}   & \textbf{SQL1}    & \textbf{SQL2}    & \textbf{SQL3}    & \textbf{SQL4}    \\
        \hline
        P("SELECT")        & \heatcolor{1.00} & \heatcolor{1.00} & \heatcolor{1.00} & \heatcolor{1.00} \\
        P("FROM")          & \heatcolor{1.00} & \heatcolor{1.00} & \heatcolor{1.00} & \heatcolor{1.00} \\
        P("WHERE")         & \heatcolor{0.98} & \heatcolor{0.53} & \heatcolor{0.51} & \heatcolor{0.44} \\
        P("JOIN")          & \heatcolor{0.00} & \heatcolor{0.13} & \heatcolor{0.21} & \heatcolor{0.18} \\
        P("GROUP")         & \heatcolor{0.01} & \heatcolor{0.11} & \heatcolor{0.20} & \heatcolor{0.26} \\
        P("ORDER")         & \heatcolor{0.98} & \heatcolor{0.39} & \heatcolor{0.12} & \heatcolor{0.30} \\
        P("HAVING")        & \heatcolor{0.01} & \heatcolor{0.06} & \heatcolor{0.01} & \heatcolor{0.12} \\
        P("UNION")         & \heatcolor{0.00} & \heatcolor{0.04} & \heatcolor{0.00} & \heatcolor{0.13} \\
        P("nested select") & \heatcolor{0.00} & \heatcolor{0.07} & \heatcolor{0.12} & \heatcolor{0.15} \\
      \end{tabular}}
  \end{subtable}
\end{table*}

\begin{itemize}[leftmargin=*,topsep=0pt]
  \item \emph{System Under Test (SUT):} We use a small PostgreSQL database consisting of three tables designed to include typical relational constructs (primary/foreign keys, nullable columns, and numeric/text fields). Queries are executed against PostgreSQL, and any runtime error or anomalous behavior is recorded. We mark a small set of fields as \emph{sensitive} (e.g., \texttt{ssn\_number}, \texttt{email}) to emulate potential data-exfiltration targets; these choices are arbitrary but fixed across experiments for comparability. This setup serves as a bug oracle by detecting whether generated queries access or attempt to exfiltrate data from the sensitive columns.

  \item \emph{Bug Oracles:} We define a set of synthetic SQL bug oracles. A query satisfies an oracle if it contains all required SQL tokens in a specified order (e.g., \texttt{JOIN}, \texttt{GROUP BY}, \texttt{DISTINCT}, \texttt{ORDER BY}) and references the sensitive columns. The oracles vary in complexity, ranging from simple single-token patterns to combinations of multiple operators or nested subqueries. These bug oracle predicates are inspired by \emph{injection-style attacks} that manipulate query structure to expose sensitive data, as well as \emph{planner, optimizer, or semantic regression bugs} triggered by specific combinations of SQL constructs. Figure~\ref{fig:heatmap-general-sql} lists 15 such bug oracles (BUG01-BUG15).
  \item \emph{Seed Sets:} For SQL, we create four main seed sets\textemdash \emph{Homogeneous-Structure}, \emph{Homogeneous-Column}, \emph{Very Small}, and \emph{Diverse (High-Coverage)}\textemdash which vary in size, structural diversity, and coverage of sensitive columns. Table~\ref{tab:seed_sets} summarizes the characteristics of these seed sets and reports marginal probabilities of key SQL constructs for representative seeds. Crucially, none of these seed sets trigger the aforementioned bug oracles; this emulates a realistic scenario where testers aim to discover new, unknown bugs by initializing the fuzzer with benign seed examples.
\end{itemize}
\vspace{2em}

\subparagraph{\textbf{XML testbed}.}

\begin{table*}[t]
  \centering
  \caption{(left) Characteristics of the XML seed sets used in our evaluation. (right) Marginal probability of key XML constructs across four subdomains.}
  \label{tab:xml_seed_sets}
  \begin{subtable}[t]{0.58\textwidth}
    \centering
    \resizebox{\linewidth}{!}{%
      \begin{tabular}{l >{\centering\arraybackslash}p{3 cm} > {\centering\arraybackslash}p{1cm} >{\centering\arraybackslash}p{4cm} >{\centering\arraybackslash}p{5cm}}
        \toprule
        \textbf{ID} & \textbf{Seed Set}       & \textbf{Size} & \textbf{Main Property}                                                 & \textbf{Expected Effect}                                            \\
        \midrule
        XML1        & Homogeneous Structure   & 15            & Single self-closing tags, varying tag types and attributes             & Low structural depth, low chance of exposing CDATA or comments bugs \\
        \midrule
        XML2        & Nested Structure        & 15            & Deeply nested `$<$request$>$` or `$<$file$>$` elements                 & Higher chance to expose XPath or CDATA bugs                         \\
        \midrule
        XML3        & Comment Heavy           & 15            & Frequent inline comments, including `$<$user$>$`/`$<$role$>$` mentions & Higher chance to expose comments related bugs                       \\
        \midrule
        XML4        & Diverse (High-Coverage) & 50            & Mixture of all above                                                   & Maximizes coverage across structure and bug classes                 \\
        \bottomrule
      \end{tabular}}
  \end{subtable}
  \hfill
  \begin{subtable}[t]{0.41\textwidth}
    \centering
    \resizebox*{\linewidth}{!}{%
      \begin{tabular}{lcccc}
        \textbf{Literal}      & \textbf{XML1}    & \textbf{XML2}    & \textbf{XML3}    & \textbf{XML4}    \\
        \hline
        P("Name")             & \heatcolor{1.00} & \heatcolor{1.00} & \heatcolor{1.00} & \heatcolor{1.00} \\
        P("CDATA")            & \heatcolor{0.00} & \heatcolor{0.33} & \heatcolor{0.01} & \heatcolor{0.17} \\
        P("COMMENT")          & \heatcolor{0.00} & \heatcolor{0.01} & \heatcolor{0.95} & \heatcolor{0.17} \\
        P("EntityRef")        & \heatcolor{0.00} & \heatcolor{0.00} & \heatcolor{0.00} & \heatcolor{0.20} \\
        P("nested structure") & \heatcolor{0.02} & \heatcolor{0.96} & \heatcolor{0.35} & \heatcolor{0.47} \\
        \multicolumn{5}{c}{
          \makecell[l]{
        * The last row, `nested structure`, is computed as                                                \\
        \qquad$P(\text{SLASH})\cdot P(\text{SLASH} \mid \text{SLASH})$                                    \\
        \qquad$ + P(\text{SLASH\_CLOSE}) \cdot P(\text{SLASH} \mid \text{SLASH\_CLOSE})$,                 \\
        \quad where \texttt{SLASH} and \texttt{SLASH\_CLOSE} corresponds to `$/$`                         \\
            \quad and `$/>$`, providing a proxy for nested XML elements.
          }
        }                                                                                                 \\
      \end{tabular}}
  \end{subtable}
\end{table*}

\begin{itemize}[leftmargin=*,topsep=0pt]
  \item \emph{System Under Test (SUT):} We create a synthetic XML project (\texttt{auth\_service}) with configuration files and data fixtures. Each XML input is executed by a safe harness that parses the document, evaluates XPath expressions, and processes application fields. The harness flags exceptions, malformed XPath, unexpected control-flow paths, and unsafe file paths. We mark a small set of files (e.g., \texttt{session\_tokens.txt}, \texttt{.env}) as \emph{sensitive} to emulate file-access oracles.
  \item \emph{Bug Oracles:} XML predicates cover both static patterns (e.g., CDATA or comment placement) and dynamic behaviors (e.g., file reads, XPath evaluation, arithmetic errors), capturing realistic XML-processing failure modes \cite{CVE-2018-14665,CWE-643,CWE-190}. An input matches an oracle only if the static condition (if any) is satisfied and the corresponding dynamic behavior is triggered; for example, a $<$file$>$ element with a path attribute may lead to a sensitive file read, while carefully placed comments can bypass logic. We also include a simple \textit{control\_bug} (an XPath inside a $<$query$>$ element that does not begin with `/') to provide an easily reachable baseline for \Grammarinator.
  \item \emph{Seed Sets:} We design four XML seed types—\emph{Homogeneous-Structure}, \emph{Nested-Request}, \emph{Comment-Heavy}, and \emph{Diverse (High-Coverage)}—to vary structural depth and token coverage. Table~\ref{tab:xml_seed_sets} summarizes their characteristics and
        reports marginal token probabilities.
\end{itemize}

%% file: chapters/4-rq2-conditioning.tex
\subsection{RQ2 (Impact of Targeted Sampling via Conditioned Generation)}
\paragraph{Methodology.}

For both SQL and XML domains, we analyze the effect of conditioning with a specific token to assess whether it improves bug coverage and input diversity. This is to emulate the scenario where a user aims to generate concrete test inputs with a specific hypothesis (i.e., a constraint). We generate 10{,}000 inputs three times, compute the number of distinct inputs triggering each bug, and average the results across runs. We compare these results against the \ExplainFuzz baseline, sampling without conditioning.
We report the following metrics:
\begin{itemize}[leftmargin=*]
  \item \textbf{Alignment}: fraction of bugs triggered by the conditioned model. For each conditioning token, we identify the bugs whose triggering conditions are semantically associated with that token (e.g., \texttt{CDATA}-related bugs for the \texttt{CDATA} token). Alignment measures whether conditioning successfully steers generation toward these target bugs.
  \item \textbf{Bug coverage}: percentage of all bugs in the testbed triggered by at least one generated input. This measures the breadth of fault detection regardless of which specific token was used for conditioning.
  \item \textbf{New bugs}: bugs triggered by the conditioned model that the unconditioned baseline fails to reach. This quantifies the incremental fault-finding benefit of targeted conditioning.
  \item \textbf{Diversity}: for bug $i$ and model $M$, let $D_M(i)$ be the number of distinct inputs (out of 10\,000) that trigger bug $i$. We report the per-bug difference $\Delta_{M,B}(i)=D_M(i)-D_B(i)$ relative to baseline $B$, and define average diversity as $\mu(\text{Div.})=\frac{1}{|B|}\sum_{i} \Delta_{M,B}(i)$. Positive values indicate that model $M$ produces more structurally varied bug-triggering inputs than the baseline.
\end{itemize}

To assess the impact of token-level conditioning on bug discovery and input diversity, we conducted a detailed analysis for both SQL and XML domains. Here, we present representative results of the XML1 and SQL2 seed sets along with the global summary due to space constraints; additional results for SQL and other XML seed sets are provided in the artifacts (Section~\ref{sec:data-avail}).

\begin{figure*}
  \centering
  \begin{subfigure}[t]{0.4\textwidth}
    \centering
    \vspace{0pt}
    \includegraphics[width=\textwidth,clip,trim=0 10 0 0]{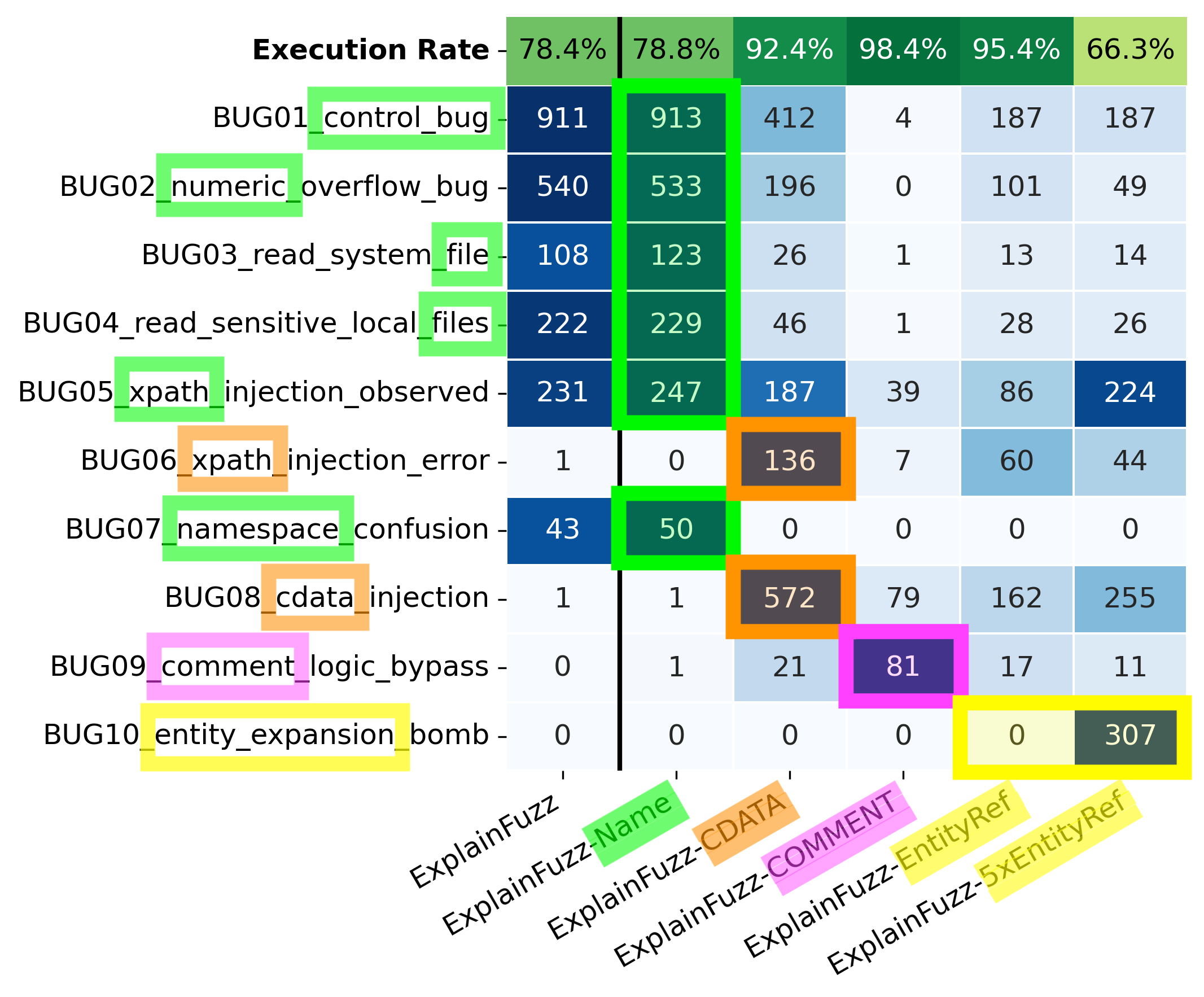}
    \caption{Bug-level heatmap showing the number of distinct queries triggering each bug for each model.}
    \label{fig:bug-level-heatm-XML1}
  \end{subfigure}
  \hfill
  \begin{subfigure}[t]{0.59\textwidth}
    \centering
    \vspace{0pt}
    \small
    \resizebox{\textwidth}{!}{
      \begin{tabular}{ccccc}
        \toprule
        \multirow{2}{*}{\makecell{Conditioning                                                        \\ Token}} & \multirow{2}{*}{Align.} & \multirow{2}{*}{\makecell{Bug                                                                     \\ Coverage}}              & \multicolumn{2}{c}{vs Unconditioned} \\
        \cmidrule(lr){4-5}
                                 &               &                  & \#NewB     & $\mu(\text{Div.})$ \\
        \midrule
        Name                     & 5/6           & 66.7\%           & 0          & +6.56              \\
        COMMENT                  & 1/1           & 53.3\%           & 1          & +80.67             \\
        CDATA                    & 2/2           & 73.3\%           & 0          & +353.33            \\
        EntityRef                & 0/1           & 63.3\%           & 0          & 0.00               \\
        5xEntityRef              & 1/1           & 90.0\%           & 1          & +307.00            \\
        \midrule
        \textbf{Global result  } & \textbf{9/11} & \textbf{100.0\%} & \textbf{2} & \textbf{+149.51}   \\
        \bottomrule
      \end{tabular}}
    \caption{Summary of statistics. \#NewB denotes the number of new bugs discovered. $\mu(\text{Div.})$ indicates the average per-bug difference in the number of distinct bug-triggering inputs (out of 10\,000) produced by the conditioned model compared to the baseline.}
    \label{tab:xml1_summary}
  \end{subfigure}
  \caption{Effect of conditioning on \ExplainFuzz for the XML1 seed set.
    The heatmap (left) shows per-bug distinct triggers, while the table (right) summarizes alignment, coverage, and diversity improvements.}
  \label{fig:xml1_cond_overview}
\end{figure*}

\paragraph{Results}

Figure~\ref{fig:bug-level-heatm-XML1} presents the heatmap for XML1, while Table~\ref{tab:xml1_summary} summarizes the conditioned generation results. Each row represents a bug, each cell in the heatmap shows the number of distinct inputs triggering a given bug, and columns 3 through 7 represent five different conditioning cases. For example, \ExplainFuzz-\hlcolor{orange!30}\texttt{CDATA} represents test generation that requires a token \hlcolor{orange!30}{\texttt{CDATA}} in the XML grammatical structure. Colored rectangles denote the expected alignment between the conditioning token and its corresponding bugs.

Conditioning improves bug discovery and diversity: across configurations, \ExplainFuzz with conditioning achieves full bug coverage (100\%) and triggers 2 new bugs compared to unconditioned \ExplainFuzz.
Conditioning with structural tokens such as \texttt{CDATA} and \texttt{5$\times$EntityRef} yields the strongest diversity gains, with average improvements of $+149$ and up to $+353$ distinct inputs.
The alignment metric shows that 9 out of 11 conditioned tokens correctly steer generation toward their expected bugs, demonstrating the effectiveness of token-guided conditioning in focusing on relevant structural regions.

Examining each conditioning token separately reveals uneven effects:
conditioning on \texttt{Name} adds little because most inputs already contain it, while \texttt{CDATA} and \texttt{COMMENT} meaningfully boost diversity and can surface additional bugs.
The strongest gain comes from compositional conditioning: only \texttt{5$\times$EntityRef} triggers \texttt{BUG10\_entity\_expansion\_bomb} (307 distinct failures), showing that multi-token constraints can expose deep structural faults that single-token conditioning misses, even with a slightly lower execution rate.

For SQL2 (figure and table omitted for space; see artifacts), token-specific conditioning similarly improves bug discovery and diversity. Tokens such as \texttt{ORDER} and \texttt{WHERE} drive notable diversity increases across their target bugs. Compared to unconditioned \ExplainFuzz, the conditioned models trigger 5 new bugs with an average diversity improvement of $+45.25$.
Two-thirds (10/15) of conditioned tokens align with their expected bugs, confirming that token-guided conditioning effectively steers generation toward relevant structural regions. Conditioning on \texttt{GROUP} discovers 3 new bugs despite a low execution rate (4.2\%), while \texttt{ORDER} yields strong diversity gains ($+274$ and $+32$ over unconditioned \ExplainFuzz). \texttt{UNION} conditioning has the lowest execution rate (0.4\%) due to structural complexity, yet still triggers one additional bug.

\begin{table*}
  \centering
  \caption{Global summary of conditioning evaluation for XML and SQL domain. $\Delta_{\text{Cov}}$ shows bug coverage improvement from conditioning; $\Delta_{\text{Div}}$ shows the increase in distinct bug-triggering inputs (out of 10,000) due to conditioning.}
  \label{tab:global_summary_cond}
  \resizebox{.5\textwidth}{!}{%
    \begin{tabular}{l ccc c}
      \toprule
                      & \multicolumn{3}{c}{\textbf{Bug Coverage (\%)}} & \textbf{Diversity}                                                 \\
      \cmidrule(lr){2-4} \cmidrule(lr){5-5}
      Seed            & \ExplainFuzz                                   & EF+Cond            & $\Delta_{\text{Cov}}$ & $\Delta_{\text{Div}}$ \\
      \midrule
      SQL1            & 17.6\%                                         & 33.3\%             & +15.7\%               & --3.46                \\
      SQL2            & 43.1\%                                         & 86.7\%             & +43.6\%               & +45.25                \\
      SQL3            & 47.1\%                                         & 66.7\%             & +19.6\%               & +19.85                \\
      SQL4            & 51.0\%                                         & 66.7\%             & +15.7\%               & +22.67                \\
      \midrule
      \textbf{Global} & \textbf{39.7}\%                                & \textbf{63.3}\%    & \textbf{+23.6}\%      & \textbf{+21.08}       \\
      \midrule
      XML1            & 66.7\%                                         & 100.0\%            & +33.3\%               & +149.51               \\
      XML2            & 86.7\%                                         & 100.0\%            & +13.3\%               & +124.49               \\
      XML3            & 86.7\%                                         & 100.0\%            & +13.3\%               & +98.11                \\
      XML4            & 90.0\%                                         & 100.0\%            & +10.0\%               & +113.65               \\
      \midrule
      \textbf{Global} & \textbf{82.5}\%                                & \textbf{100.0}\%   & \textbf{+17.5}\%      & \textbf{+121.44}      \\
      \bottomrule
    \end{tabular}}
\end{table*}

Table~\ref{tab:global_summary_cond} summarizes conditioning results across all seeds for SQL and XML. In SQL, average bug coverage rises from 39.7\% to 63.3\% (+23.6\%), and input diversity increases by +21.08 distinct bug-triggering inputs on average. In XML, conditioning lifts coverage from 82.5\% to 100\% (+17.5\%) and boosts diversity by +121.44 distinct inputs. Conditioning therefore not only aligns generation with bug-relevant structures but also substantially increases the variety of inputs that trigger each bug, enabling more comprehensive test coverage.

Seed design also matters. Richer seeds (e.g., SQL4, XML4) yield higher coverage and diversity overall, while homogeneous seeds (XML1) gain the most from conditioning (largest relative diversity increase, $+149.51$). This suggests conditioning compensates for limited seed variety, and when seeds are already diverse it acts as fine-grained guidance toward bug-relevant constructs.

\paragraph{Key Insights.}

The conditioning evaluation clarifies the specific contribution of the PC within \ExplainFuzz. It shows that the PC captures structural correlations that can be actively exploited, revealing latent co-occurrence patterns (e.g., around \texttt{JOIN}, \texttt{WHERE}, or \texttt{CDATA}) that only become usable under conditioning. Conditioning is especially valuable with sparse or simple seeds, where it amplifies underrepresented patterns while preserving semantic coherence and avoiding degenerate templates.

\paragraph{\textbf{RQ2 Result Summary.}}
Probabilistic conditioning substantially improves both bug coverage (+23.6\% for SQL, +17.5\% for XML) and input diversity (+21 distinct triggers per bug for SQL, +121 for XML) by steering generation toward token-aligned structures that static grammar-based generation misses. High alignment rates (9/11 for XML, 10/15 for SQL) confirm that conditioning effectively targets bug-relevant constructs while generating structurally varied inputs.

%% file: chapters/4-rq3-bugfinding.tex
\subsection{RQ3 (Bug Triggering Input Diversity)}
\label{sec:rq3_bug_diversity}

\paragraph{Methodology.}

We first use \Grammarinator to generate three corpora of 10,000 inputs each from the chosen seed set; we evaluate across multiple seed sets per domain. \ExplainFuzz is then trained on one of these (anonymized) corpora and used to generate three corpora of 10,000 inputs, mirroring the same data volume and number of trials as \Grammarinator. All generated inputs are executed on the system under test (SUT), and we report the average across the three independent corpora for each tool.
We measure the following metrics:
\begin{itemize}[leftmargin=*]
  \item \textbf{Executable rate}: the percent of generated inputs successfully parsed and executed.
  \item \textbf{Bug coverage}: the percent of all synthetic bugs that were triggered.
  \item \textbf{Total triggers}: the total number of inputs triggering any bug (including duplicates).
  \item \textbf{Distinct triggering inputs}: the number of unique inputs that triggered at least one bug.
  \item \textbf{Explicit sensitive} (SQL only): the percent of distinct triggering inputs that mention at least one sensitive field and do not use `SELECT *'.
\end{itemize}
In addition, we perform a \emph{qualitative assessment} of a representative subset of bug-triggering inputs, focusing on syntactic validity, semantic plausibility, and structural diversity.

We first report aggregate performance averaged over the three runs of 10,000 generated inputs each, followed by detailed per-bug coverage analysis. Finally, we qualitatively examine the quality of generated bug-triggering inputs.

\subsubsection{Aggregate Bug-Finding Performance}

\begin{table*}[t]
  \centering
  \caption{Multi-bug evaluation per seed set: \Grammarinator vs \ExplainFuzz}
  \resizebox{\textwidth}{!}{%
    \begin{tabular}{cc|rrrrr}
      \toprule
      \multirow{2}{*}{\textbf{\makecell{Seeds                                                             \\ (Description)}}} & \multirow{2}{*}{\textbf{\makecell{Model}}} & \multirow{2}{*}{\textbf{\makecell{Executable \\ rate (\%)}}} & \multirow{2}{*}{\textbf{\makecell{Bug \\ coverage (\%)}}} & \multirow{2}{*}{\textbf{\makecell{Total \\ triggers}}} & \multirow{2}{*}{\textbf{\makecell{Distinct \\ triggering inputs}}} & \multirow{2}{*}{\textbf{\makecell{Explicit \\ sensitivity (\%)}}} \\
       &              &               &               &                 &                 &               \\
      \midrule
      \multirow{2}{*}{\makecell{SQL1                                                                      \\ (Homogeneous structure)}} & \Grammarinator                     & 45.1         & \textbf{20.0} & 132.0         & 27.0          & 0.0           \\
       & \ExplainFuzz & \textbf{73.2} & \textbf{20.0} & \textbf{1060.0} & \textbf{1057.0} & \textbf{78.6} \\
      \multirow{2}{*}{\makecell{SQL2                                                                      \\ (Homogeneous column)}} & \Grammarinator                        & 43.1         & \textbf{53.3} & 80.3          & 30.3          & 0.0           \\
       & \ExplainFuzz & \textbf{50.0} & 48.9          & \textbf{426.3}  & \textbf{309.7}  & \textbf{65.0} \\
      \multirow{2}{*}{\makecell{SQL3                                                                      \\ (Small set + diverse)}} & \Grammarinator                     & 40.0         & 40.0          & 123.7         & 10.0          & 0.0           \\
       & \ExplainFuzz & \textbf{43.4} & \textbf{53.3} & \textbf{212.7}  & \textbf{168.7}  & \textbf{62.5} \\
      \multirow{2}{*}{\makecell{SQL4                                                                      \\ (Big set + diverse)}} & \Grammarinator                        & 35.9         & 53.3          & 75.0          & 25.0          & 0.0           \\
       & \ExplainFuzz & \textbf{36.1} & \textbf{57.8} & \textbf{310.7}  & \textbf{224.7}  & \textbf{62.9} \\
      \midrule
      \multirow{2}{*}{\makecell{XML1                                                                      \\ (Homogeneous structure)}} & \Grammarinator                     & 38.5         & 10.0          & 69.7          & 68.0          & --            \\
       & \ExplainFuzz & \textbf{78.4} & \textbf{66.7} & \textbf{2055.7} & \textbf{2054.3} & --            \\
      \multirow{2}{*}{\makecell{XML2                                                                      \\ (Nested structure)}} & \Grammarinator                        & \textbf{56.3} & 10.0          & 470.0         & 88.0          & --            \\
       & \ExplainFuzz & 43.0          & \textbf{86.7} & \textbf{1428.7} & \textbf{1428.3} & --            \\
      \multirow{2}{*}{\makecell{XML3                                                                      \\ (Comments heavy)}} & \Grammarinator                      & 59.9         & 10.0          & 234.0         & 41.3          & --            \\
       & \ExplainFuzz & \textbf{81.3} & \textbf{86.7} & \textbf{1894.0} & \textbf{1893.3} & --            \\
      \multirow{2}{*}{\makecell{XML4                                                                      \\ (Big set + diverse)}} & \Grammarinator                        & 56.5         & 10.0          & 179.3         & 34.0          & --            \\
       & \ExplainFuzz & \textbf{79.2} & \textbf{90.0} & \textbf{1223.3} & \textbf{1204.0} & --            \\
      \bottomrule
    \end{tabular}}
  \label{tab:global-sql-xml}
\end{table*}

Table~\ref{tab:global-sql-xml} summarizes the multi-bug evaluation across all SQL and XML seed sets. Across both domains, \ExplainFuzz achieves higher executable rates, broader bug coverage, and—most distinctly—far more \emph{unique} bug-triggering inputs. The magnitude of \ExplainFuzz's improvements over \Grammarinator suggests that the PC's ability to capture structural patterns from seed inputs, combined with the domain-specific concretizers, plays a significant role in enabling semantically meaningful inputs that reach deeper program behaviors and trigger more bugs. Below, we analyze results per domain.

\paragraph{{SQL domain.}} Across all SQL seed sets, execution rates increase modestly (+2--28\%), and overall bug coverage is comparable between the two tools. The main difference appears in the \emph{volume and diversity} of triggering inputs: \ExplainFuzz produces on average \emph{+417 additional distinct} triggers per seed set. Because the PC generates a diverse set of structural patterns and the concretizer instantiates them with schema-aware variations, the system explores a wider semantic neighborhood than grammar-only mutation.

The only notable deviation is SQL2, where \Grammarinator attains slightly higher bug coverage. However, its triggers predominantly rely on generic patterns (e.g., \texttt{SELECT *}) and do not reference bug-relevant columns, reflected in its 0\% explicit-sensitive rate. The SQL concretizer in \ExplainFuzz, by contrast, fills \texttt{ID} placeholders with real table and column names drawn from schema metadata, yielding 65--79\% explicit-sensitive queries.

\paragraph{{XML domain.}} For three of the four XML seed sets (\texttt{XML1, XML3, XML4}), \ExplainFuzz achieves markedly higher executable rates (78--81\% vs.\ 38--60\%). \texttt{XML2} is the exception: deep, highly nested structures are better preserved by \Grammarinator's mutation strategy, leading to a higher execution rate in that case.
Across all XML seed sets, bug coverage strongly favors \ExplainFuzz (66--90\% vs.\ 10\%). The difference in input diversity is even more pronounced: \ExplainFuzz generates an average of \emph{+1587 additional distinct} bug-triggering documents per seed set, indicating that it explores a much broader range of XML variations than grammar-only generation. This occurs because PC samples encode varied structural patterns and the concretizer grounds them in realistic project values (file paths, entity names, boundary-like attributes); thus, the resulting XML inputs explore many more semantically relevant configurations.

\subsubsection{Per-Bug Coverage Analysis}
\begin{figure*}[htt]
  \centering
  \begin{subfigure}[b]{0.49\textwidth}
    \centering
    \includegraphics[width=\linewidth,clip,trim=0 10 0 0]{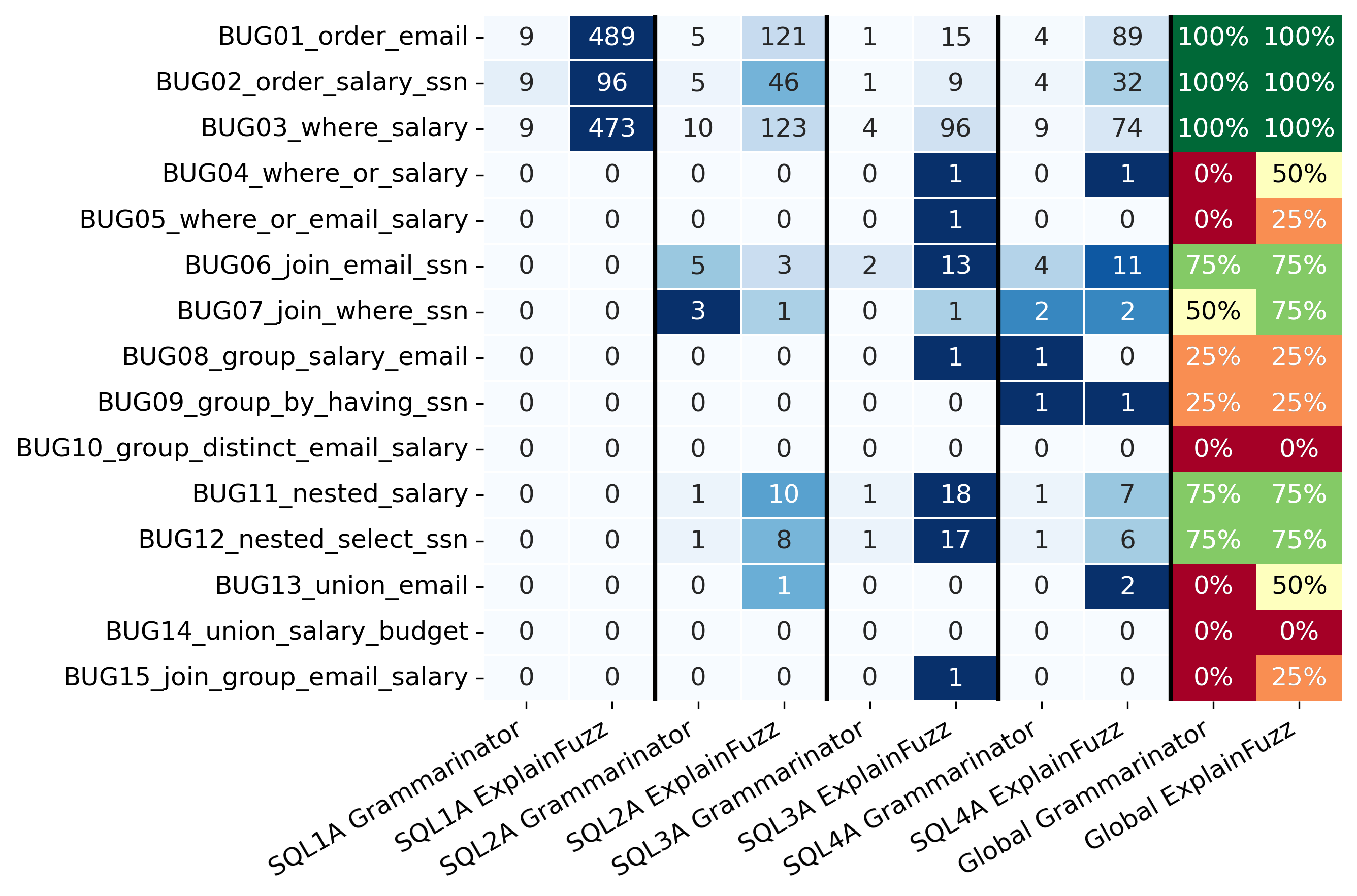}
    \caption{SQL}
    \label{fig:heatmap-general-sql}
  \end{subfigure}
  \hfill
  \begin{subfigure}[b]{0.49\textwidth}
    \centering
    \includegraphics[width=\linewidth,clip,trim=0 10 0 0]{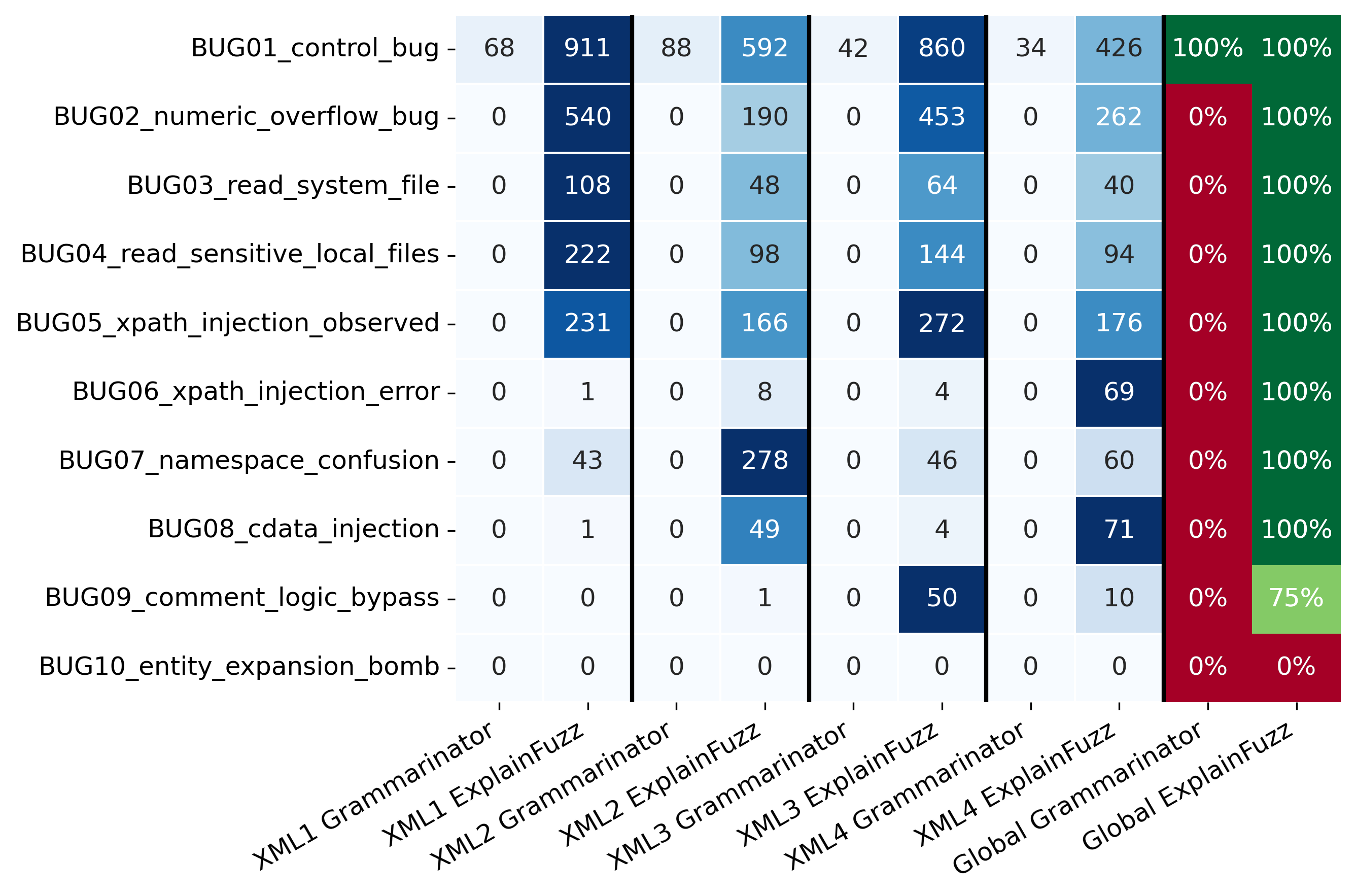}
    \caption{XML}
    \label{fig:heatmap-general-xml}
  \end{subfigure}
  \caption{Bug trigger summary across seeds and models; the x-axis shows bug IDs, the y-axis lists seed sets and models, color intensity denotes the number of distinct inputs triggering each bug, and the rightmost columns report the global success ratio per bug.}
  \label{fig:heatmap-general}
\end{figure*}

For each synthetic bug predicate, we record the number of \textit{distinct inputs} that trigger it (averaged over three runs per seed set) and compute the \textit{global success ratio}, defined as the fraction of seed-trained models that trigger each bug. A bug is considered \textit{triggered} if at least one input generated in any run causes it to fail. Figures~\ref{fig:heatmap-general-sql} and~\ref{fig:heatmap-general-xml} visualize these per-bug results for SQL and XML; darker cells indicate more distinct triggers, and the rightmost columns report the global success ratio.

\paragraph{SQL domain.}
Figure~\ref{fig:heatmap-general-sql} highlights a strong and consistent advantage for \ExplainFuzz across nearly all synthetic bug predicates.
While \Grammarinator occasionally triggers low-complexity bugs (e.g., \texttt{BUG01\_order\_email}, \texttt{BUG02\_order\_salary\_ssn}, \texttt{BUG03\_where\_salary}), it does so with only a handful of distinct inputs (1--9 per seed set).
In contrast, \ExplainFuzz generates hundreds of valid yet diverse inputs for the same bugs (e.g., 489, 95, and 472 distinct triggers on SQL1 alone), reflecting its ability to meaningfully mutate both projection and filtering clauses. This difference arises because \ExplainFuzz's SQL concretizer instantiates schema-aware identifiers and type-correct literals over structurally valid PC templates, enabling broader exploration of semantically relevant queries.

\ExplainFuzz also uncovers complex bugs involving multi-clause conditions such as joins and nested queries (\texttt{BUG06\_join\_email\_ssn}, \texttt{BUG11\_nested\_salary}, \texttt{BUG12\_nested\_select\_ssn}), which \Grammarinator almost never reaches.
These cases demonstrate that \ExplainFuzz can synthesize realistic cross-table dependencies and subquery patterns, enabled by its custom concretizer, which grammar-only generation cannot model.
Bugs relying on combinatorial conditions (\texttt{BUG04--BUG05}, \texttt{BUG08--BUG10}, \texttt{BUG14--BUG15}) remain mostly untriggered by both systems, suggesting that such multi-operator compositions are statistically rare in the current sampling strategy.
Overall, \ExplainFuzz achieves complete or near-complete bug coverage (100\% for 6 of 15 bugs, 75\% for 4 others), whereas \Grammarinator remains below 50\% for most cases.

\paragraph{XML domain.}
Figure~\ref{fig:heatmap-general-xml} shows a stark divergence between \Grammarinator and \ExplainFuzz.
\Grammarinator only triggers the control bug and fails to reach any other synthetic bug, confirming that it can only generate very simple inputs that remain close to the initial seeds.
In contrast, \ExplainFuzz triggers all reachable bugs and consistently produces a much larger number of distinct inputs.

\ExplainFuzz triggers many distinct inputs for file-access bugs (\texttt{read\_system\_file}: 40--108; \texttt{read\_sensitive\_local\_files}: 94--222), especially in simpler, path-rich seed sets (max XML1, min XML4). For XPath bugs, it produces many triggers for \texttt{xpath\_injection\_observed} (166--272) and fewer for \texttt{xpath\_injection\_error} (up to 69), consistent with occasional risky CDATA payloads. Numeric overflows are frequent due to schema-driven numeric attributes, while namespace confusion is modest in most seed sets (around 50 triggers) but peaks in XML2 (278) with nested elements. Comment/CDATA bugs are harder to find, with few triggers concentrated in XML3 and XML2/XML4. Both generators fail to trigger the entity-expansion bug, which requires five or more distinct entity references and remains extremely rare under current concretization.

\paragraph{Impact of the Seeds}
Seed choice affects both generators, with a stronger effect on \Grammarinator than \ExplainFuzz. \Grammarinator remains tightly bound to the structural patterns in its initial corpus: missing constructs are rarely discovered. However, seed variation also meaningfully influences \ExplainFuzz. With minimal or structurally simple seeds (e.g., \texttt{XML1}, \texttt{SQL1}), \ExplainFuzz reaches only a limited subset of bugs; in contrast, more diverse or nested seed sets (e.g., \texttt{SQL3}, \texttt{SQL4}, \texttt{XML4}) enable it to uncover a broader range of bug-triggering structures. In the XML domain, we further observe that changing seed types does not help \Grammarinator, but \ExplainFuzz extracts new patterns and triggers additional bugs. Seed quantity affects distinct triggers more than coverage; for instance, \texttt{SQL4} slightly exceeds \texttt{SQL3} on bug coverage (57.8\% vs.\ 53.3\%) but yields over 50 more distinct triggers.

\subsubsection{Qualitative Analysis of Bug-Triggering Inputs}

\begin{table*}
  \centering
  \caption{Qualitative comparison of bug-triggering inputs between \Grammarinator and \ExplainFuzz.}
  \label{tab:qualitative_examples}

  \tiny
  \resizebox{\textwidth}{!}{%
    \begin{tabular}{p{0.19\linewidth}p{0.81\linewidth}}
      \multicolumn{2}{l}{\textbf{SQL, BUG02\_ order\_ salary\_ ssn}}                                                                                                                                                   \\
      \midrule
      \textbf{\Grammarinator}             & \texttt{\tiny SELECT \textbf{*} FROM employees WHERE department\_id = 5 \textbf{ORDER BY} hire\_date DESC;}                                                                \\
      \textbf{\ExplainFuzz (anonymized)}  & \texttt{\tiny SELECT ID , ID FROM ID WHERE ID LT Integral ORDER BY Identifier USING GT;}                                                                                   \\
      \textbf{\ExplainFuzz (concretized)} & \texttt{\tiny SELECT \textbf{salary}, \textbf{ssn\_number} FROM employees WHERE performance\_ score < 8334 \textbf{ORDER BY} ssn\_number USING >;}                         \\
      \\
      \multicolumn{2}{l}{\textbf{XML, BUG01\_ control\_ bug}}                                                                                                                                                          \\
      \midrule
      \textbf{\Grammarinator}             & \texttt{\tiny <request user\_id="106" quota="200"> <query \textbf{q="\_kQw?jN)"} limit="3"/> <file path="data/ temp.xml" name="tempfile" /> </request>}                    \\
      \textbf{\ExplainFuzz (anonymized)}  & \texttt{\tiny OPEN Name Name=STRING Name=STRING CLOSE OPEN Name Name=STRING CLOSE OPEN Name CLOSE CDATA OPEN SLASH Name CLOSE OPEN SLASH Name CLOSE OPEN SLASH Name CLOSE} \\
      \textbf{\ExplainFuzz (concretized)} & \texttt{\tiny <query \textbf{q="config/ settings.xml"} select="value"> <user quota="8769076"> <file> <![CDATA[Backup file content v1]]> </file> </user> </query>}          \\
    \end{tabular}}
\end{table*}

We conducted a qualitative analysis to assess whether \ExplainFuzz produces more natural, realistic, and interpretable inputs than \Grammarinator. We manually selected a small subset of bug-triggering inputs from both tools and compared them along three axes: (i) syntactic validity and structural coherence, (ii) semantic plausibility with respect to the domain schema or configuration patterns, and (iii) diversity and richness of structural constructs. Table~\ref{tab:qualitative_examples} shows representative examples (one per domain); more examples appear in our artifacts (Section~\ref{sec:data-avail}).

Our qualitative analysis found that \Grammarinator can generate a wide range of syntactic structures, but its lack of schema awareness often yields semantically weak inputs. In contrast, \ExplainFuzz preserves structural diversity by training on anonymized \Grammarinator samples while enforcing semantic validity via schema-aware concretization.
In SQL, this leads to realistic column-predicate combinations and coherent joins; for example, it selects meaningful column pairs (e.g., salary and ssn\_number) rather than relying on generic \texttt{SELECT *} patterns.
In XML, it yields deeper, well-typed documents with CDATA, comments, and realistic attributes, while \Grammarinator often yields random attributes and shallow nesting.
Overall, \ExplainFuzz transforms raw syntactic variety into semantically grounded, executable, and interpretable test cases.

\paragraph{\textbf{RQ3 Result Summary.}}
\ExplainFuzz consistently outperforms the grammar-based baseline in bug reachability and input diversity: bug coverage improves from 35.3\% to 39.7\% in SQL and from 10\% to 82.5\% in XML, while the number of distinct bug-triggering inputs increases by +417 (SQL) and +1587 (XML) per seed set on average. This advantage arises because \Grammarinator remains tightly bound to seed structures, whereas \ExplainFuzz's learned distribution explores a wider semantic neighborhood. The qualitative analysis confirms that \ExplainFuzz's inputs are more semantically meaningful and contextually valid, balancing structural diversity with domain-specific realism.

%% file: chapters/6-threat.tex
\section{Threats to Validity}

\paragraph{Construct Validity.}
We operationalize effectiveness using bug coverage, diversity, and alignment metrics over synthetic bugs. These constructs approximate fuzzing utility but do not fully represent real-world vulnerabilities or developer priorities. Scalability is assessed implicitly through tractability of PC compilation and training; tokenization (anonymization) and length limits improve feasibility but trade expressiveness for tractability. Concretization of tokens is domain-specific, and without a suitable concretizer the pipeline may generate coherent grammatical structures, not necessarily concrete, executable inputs.

\paragraph{Internal Validity.}
Differences in execution rates and concretizer behavior can affect coverage metrics. Results are averaged over multiple runs, as stochasticity in the test generators can introduce variance.

\paragraph{External Validity.}
Our evaluation focuses on SQL and XML, where we implemented concretizers that convert tokens to concrete inputs, and we also curated a set of bug oracles for evaluation. Results may not transfer directly to other domains with the different degree of lexical variability. Diverse seeds can capture a broader set of grammatical structures with context-sensitive dependencies, while sparse seeds can overfit the learned distribution.

%% file: chapters/7-conclusion.tex
\section{Conclusion}

This paper introduces \ExplainFuzz, a grammar-aware input generator capable of capturing context-sensitive probabilistic dependencies and demonstrates that probabilistic, grammar-aware test generation can be both interpretable and controllable. It compiles a context-free grammar into a grammar-aware probabilistic circuit (PC) and trains the PC using sample seeds. Unlike traditional seed-based mutational fuzzing that perform local neighborhood search with mutations, \ExplainFuzz provides explainability and transparency by enabling probabilistic inference queries, enabling users to examine the underlying input distribution. Furthermore, by leveraging the conditioned sampling capabilities of PCs, \ExplainFuzz can generate targeted samples by constraining the generation process under specific constraints.

\ExplainFuzz outperforms grammar-unaware PCs (PC-HMM), LLMs, and probabilistic context-free grammar (pCFG) learners in terms of perplexity, indicating superior coherence and realism. Across SQL and XML domains, \ExplainFuzz consistently outperforms the state-of-the-art grammar-based mutational fuzzer, \Grammarinator, in both input diversity and bug-triggering rates. Our results indicate that conditioned sampling sharpens the controllability of targeted test generation—especially when seeds are sparse or structurally simple—by amplifying underrepresented patterns without sacrificing the contextual probabilistic dependencies essential for generating complex, hierarchical inputs.

%% file: chapters/appendix.tex
\section{Cross-bug evaluation setup complementary data}
\label{app:additional-testbeds-data}
\subsection{SQL Database Schema}
\label{app:sql-schema}

This appendix summarizes the structure of the SQL testbed used in our evaluation.  
The database consists of three tables: \texttt{employees}, \texttt{projects}, and \texttt{departments}, together with their primary keys and foreign keys.

\subsection{Table: \texttt{employees}}
\begin{table}[H]
\centering
\begin{tabular}{lll}
\hline
\textbf{Column} & \textbf{Type} & \textbf{Nullable} \\
\hline
id                & integer & no  \\
employee\_name     & text    & no  \\
hire\_date         & date    & no  \\
salary             & numeric & yes \\
full\_time         & boolean & yes \\
performance\_score & integer & yes \\
email              & text    & yes \\
ssn\_number        & text    & yes \\
department\_id     & integer & yes \\
project\_id        & integer & yes \\
\hline
\end{tabular}

\vspace{2ex}
{\raggedright
\textbf{Primary Key:} \texttt{employees\_pkey (id)} \\
\textbf{Foreign Keys:}
\begin{itemize}
\item \texttt{department\_id} $\rightarrow$ \texttt{departments(dep\_number)}
\item \texttt{project\_id} $\rightarrow$ \texttt{projects(proj\_number)}
\end{itemize}
}
\end{table}

\subsection{Table: \texttt{projects}}
\begin{table}[H]
\centering
\begin{tabular}{lll}
\hline
\textbf{Column} & \textbf{Type} & \textbf{Nullable} \\
\hline
proj\_number  & integer & no  \\
project\_name & text    & no  \\
start\_date   & date    & yes \\
end\_date     & date    & yes \\
budget        & numeric & yes \\
dep\_id       & integer & yes \\
\hline
\end{tabular}

\vspace{2ex}
{\raggedright
\textbf{Primary Key:} \texttt{projects\_pkey (proj\_number)} \\
\textbf{Foreign Keys:}
\begin{itemize}
\item \texttt{dep\_id} $\rightarrow$ \texttt{departments(dep\_number)}
\end{itemize}
}
\end{table}

\subsection{Table: \texttt{departments}}
\begin{table}[H]
\centering
\begin{tabular}{lll}
\hline
\textbf{Column} & \textbf{Type} & \textbf{Nullable} \\
\hline
dep\_number      & integer & no  \\
department\_name & text    & no  \\
location         & text    & yes \\
dep\_budget      & numeric & yes \\
\hline
\end{tabular}

\vspace{2ex}
{\raggedright
\textbf{Primary Key:} \texttt{departments\_pkey (dep\_number)} \\
\textbf{Unique Constraints:} \texttt{department\_name (unique)} \\
\textbf{Referenced By:}
\begin{itemize}
\item \texttt{employees.department\_id}
\item \texttt{projects.dep\_id}
\end{itemize}
}
\end{table}

\subsection{XML Schema Metadata}
\label{app:xml-schema}

This appendix summarizes the structure of the XML testbed used in our evaluation.  
For each XML element, we list its attributes and subelements.  

\subsection{Element: \texttt{request}}
\begin{itemize}
\item \textbf{Attributes:} \texttt{user\_id (xs:long)}, \texttt{quota (xs:long)}, \texttt{amount (xs:int)}, \texttt{repeat (xs:int)}, \texttt{action (xs:string)}, \texttt{file\_path (xs:string)}
\item \textbf{Subelements:} \texttt{query}, \texttt{file}, \texttt{user}, \texttt{message}, \texttt{user\_id}, \texttt{quota}, \texttt{amount}, \texttt{repeat}, \texttt{action}, \texttt{file\_path}
\end{itemize}

\subsection{Element: \texttt{user}}
\begin{itemize}
\item \textbf{Attributes:} \texttt{id (xs:long)}, \texttt{quota (xs:long)}, \texttt{username (xs:string)}, \texttt{role (xs:string)}
\item \textbf{Subelements:} \texttt{query}, \texttt{file}, \texttt{message}, \texttt{id}, \texttt{username}, \texttt{role}, \texttt{quota}
\end{itemize}

\subsection{Element: \texttt{file}}
\begin{itemize}
\item \textbf{Attributes:} \texttt{path (xs:anyURI)}, \texttt{name (xs:string)}, \texttt{content (xs:string)}, \texttt{author (xs:string)}, \texttt{price (xs:float)}
\item \textbf{Subelements:} \texttt{message}, \texttt{path}, \texttt{name}, \texttt{content}, \texttt{author}, \texttt{price}
\end{itemize}

\subsection{Element: \texttt{message}}
\begin{itemize}
\item \textbf{Attributes:} \texttt{from (xs:string)}, \texttt{to (xs:string)}, \texttt{content (xs:string)}
\item \textbf{Subelements:} \texttt{file}, \texttt{from}, \texttt{to}, \texttt{content}
\end{itemize}

\subsection{Element: \texttt{query}}
\begin{itemize}
\item \textbf{Attributes:} \texttt{select (xs:string)}, \texttt{q (xs:string)}, \texttt{limit (xs:int)}
\item \textbf{Subelements:} \texttt{user}, \texttt{file}, \texttt{message}, \texttt{select}, \texttt{q}, \texttt{limit}
\end{itemize}

\newpage

\subsection{XML Project Layout}

The following shows the directory structure of the \texttt{auth\_service\_xml} project used in our evaluation:

\begin{figure}[h]
    \centering
    \includegraphics[width=\textwidth]{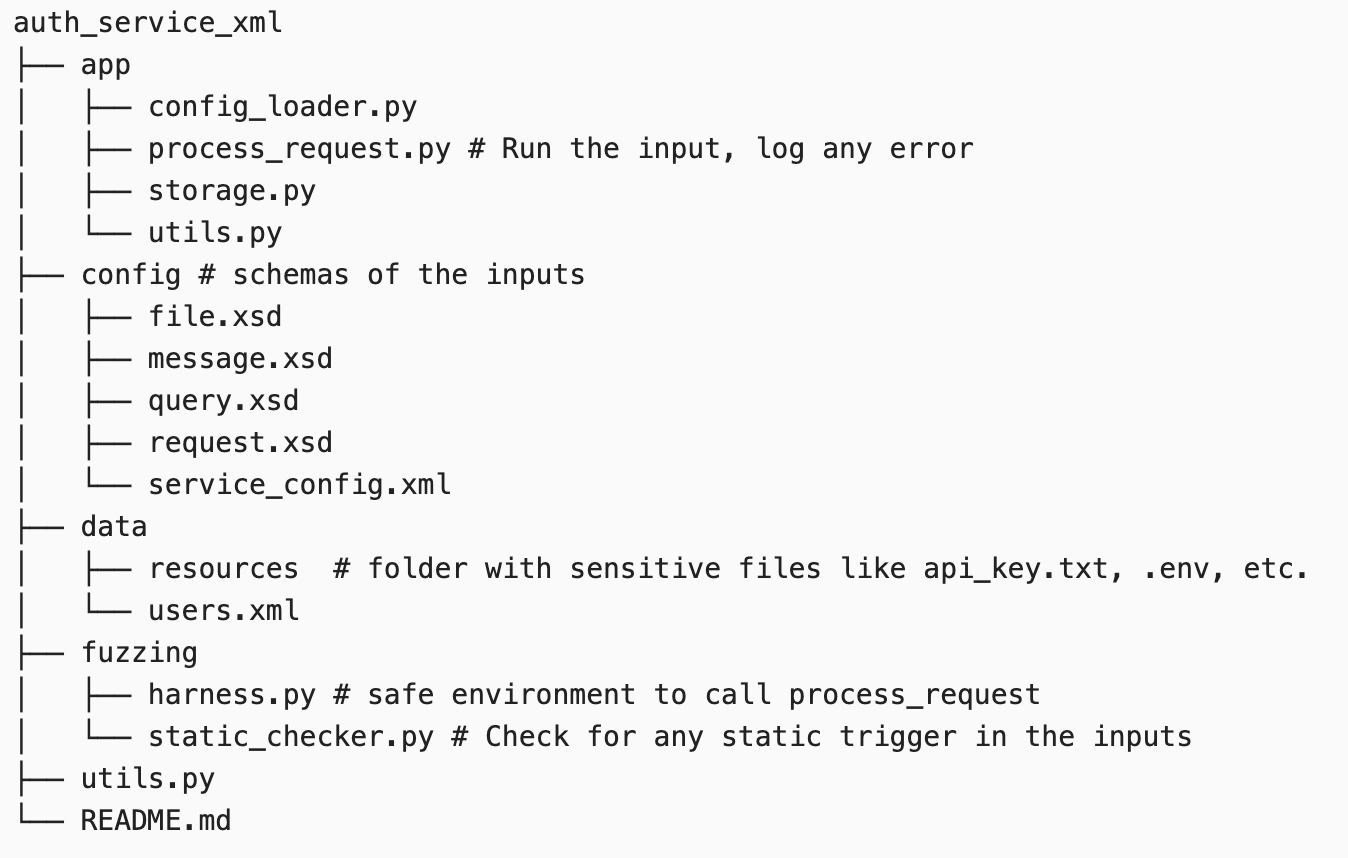}
    \caption{Layout of the XML project used as a testbed}
    \label{fig:XML-layout}
\end{figure}

\newpage

\subsection{Bug Predicates}
\label{app:bug-predicates-table}

\begin{table}[h!]
\centering
\caption{Bug predicates for XML fuzzing}
\label{tab:bug-predicates-xml}
\resizebox{.8\textwidth}{!}{%
\begin{tabular}{p{3.2cm}|p{3cm}|p{3cm}|p{6cm}}
\toprule
\textbf{Bug ID} & \textbf{Static Trigger} & \textbf{Dynamic Trigger} & \textbf{Motivation} \\
\toprule
BUG01\_control \_bug & incorrect\_start \_query & --- & Detect if an xpath inside a query element doesn't start with '/'; serves as a reachability baseline. \\
\midrule
BUG02\_numeric \_overflow\_bug & numeric\_overflow \_candidate & --- & Detect numeric overflows causing logic errors or potential DoS. \\
\midrule
BUG03\_read \_system\_file & --- & system\_file\_read & Identify dynamic reads that expose sensitive files via XXE or path traversal. \\
\midrule
BUG04\_read \_sensitive\_local \_files & --- & local\_repo\_file \_read & Detect reads of local sensitive files, indicating possible information leakage. \\
\midrule
BUG05\_xpath \_injection\_observed & xpath\_injection \_candidate & xpath\_query \_executed & Execution of untrusted XPath queries shows potential data access or auth bypass. \\
\midrule
BUG06\_xpath \_injection\_error & xpath\_injection \_candidate & xpath\_injection \_error & Errors from attacker-controlled XPath indicate validation flaws and DoS vectors. \\
\midrule
BUG07\_namespace \_confusion & namespace \_confusion \_candidate & --- & Confusing namespaces may bypass schema validation or access controls. \\
\midrule
BUG08\_cdata \_injection & cdata\_injection \_candidate & --- & Embed risky CDATA content to test unsafe downstream processing. \\
\midrule
BUG09\_comment \_logic\_bypass & auth\_bypass \_detected & --- & Comments in critical positions can alter behavior or bypass logic. \\
\midrule
BUG10\_entity \_expansion\_bomb & entity\_expansion \_critic & --- & Generate many entity references to test parser resilience and DoS. \\
\bottomrule
\end{tabular}}
\end{table}

\begin{table}[h!]
\centering
\caption{SQL Bug Predicates}
\label{tab:sql-bug-predicates}
\footnotesize 
\resizebox{.8\textwidth}{!}{%
\begin{tabular}{p{3.5cm}|p{1.9cm}|p{2.2cm}|p{7.3cm}}
\toprule
\textbf{Bug ID} & \textbf{Token Required} & \textbf{Sensitive Columns} & \textbf{Motivation} \\
\toprule
BUG01\_order \_email & ORDER, BY & email & Ordering on long text fields may expose encoding, collation, or comparator bugs. \\
\midrule
BUG02\_order \_salary\_ssn & ORDER, BY & salary, ssn\_number & ORDER BY on mixed/large types can trigger correctness and stability bugs. \\
\midrule
BUG03\_where \_salary & WHERE & salary & Type coercion in WHERE may cause incorrect filtering or planner errors. \\
\midrule
BUG04\_where\_or \_salary & WHERE, OR & salary & Short-circuit/NULL-handling inconsistencies can lead to wrong row selection. \\
\midrule
BUG05\_where\_or \_email\_salary & WHERE, OR & email, salary & Compound OR on sensitive fields may trigger logic simplification bugs. \\
\midrule
BUG06\_join\_email \_ssn & JOIN & email, ssn\_number & Join reorderings may expose protected columns or bypass filters. \\
\midrule
BUG07\_join\_where \_ssn & JOIN, WHERE & ssn\_number & Planner reorderings with WHERE can violate row-level security. \\
\midrule
BUG08\_group \_salary\_email & GROUP, BY & salary, email & Aggregation on sensitive fields may trigger overflow, rounding, or mis-grouping errors. \\
\midrule
BUG09\_group\_by \_having\_ssn & GROUP, BY, HAVING & ssn\_number & HAVING mis-evaluation can produce incorrect grouped row inclusion/exclusion. \\
\midrule
BUG10\_group \_distinct\_email\_salary & GROUP, BY, DISTINCT & email, salary & DISTINCT + GROUP BY can yield deduplication or aggregation corruption. \\
\midrule
BUG11\_nested \_salary & SELECT, FROM, SELECT & salary & Nested subqueries with aggregation can cause crashes or wrong joins. \\
\midrule
BUG12\_nested \_select\_ssn & SELECT, FROM, SELECT & ssn\_number & Deeply nested SELECTs can expose semantic or permission propagation errors. \\
\midrule
BUG13\_union \_email & UNION & email & UNION-based injection can exfiltrate sensitive fields via projection mismatches. \\
\midrule
BUG14\_union \_salary\_budget & UNION & salary, budget & UNION on mismatched numeric types may cause coercion, overflow, or schema bugs. \\
\midrule
BUG15\_join\_group \_email\_salary & JOIN, GROUP, BY & salary, email & Join + aggregation interactions may produce planner errors or wrong visibility. \\
\bottomrule
\end{tabular}}
\end{table}

\newpage

\section{Additional evaluation conditioning results}
\label{app:additional-cond-results}

\begin{figure*}[ht]
    \centering
    \begin{subfigure}[t]{0.95\textwidth}
        \centering
        \includegraphics[width=\textwidth]{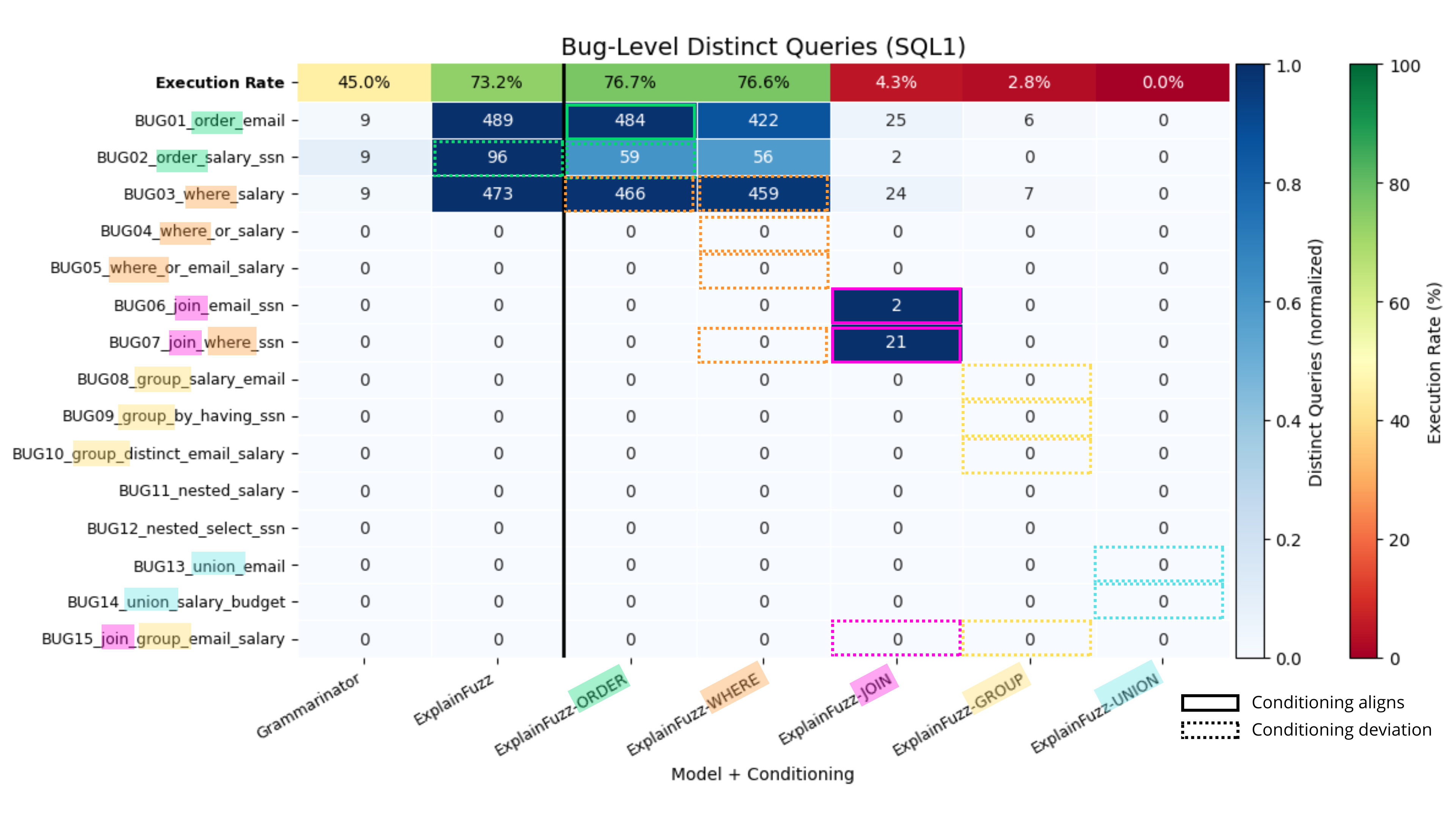}
        \caption{Bug-level heatmap showing the number of distinct queries triggering each bug for each model.}
        \label{fig:bug-level-heatm-SQL1}
    \end{subfigure}
    \vspace{0.5em}
    \begin{subfigure}[t]{0.95\textwidth}
        \centering
        \small
        \resizebox{\textwidth}{!}{
        \begin{tabular}{l >{\centering\arraybackslash}p{1.5cm} > {\centering\arraybackslash}p{1.6cm} >{\centering\arraybackslash}p{1.2cm} >{\centering\arraybackslash}p{1.8cm} >{\centering\arraybackslash}p{1.2cm} >{\centering\arraybackslash}p{1.8cm}}
        \toprule
        & & & \multicolumn{2}{c}{vs \ExplainFuzz} & \multicolumn{2}{c}{vs \Grammarinator} \\
        \cmidrule(lr){4-5} \cmidrule(lr){6-7}
        Conditioned Model & Alignment & Bug Coverage & New Bugs & Avg Diversity & New Bugs & Avg Diversity \\
        \midrule
        \ExplainFuzz + ORDER & 0/2 & 17.6\%  & 0 & -21.33 & 0 & +262.00 \\
        \ExplainFuzz + WHERE & 0/4 & 17.6\%  & 0 & -3.42 & 0 & +112.42 \\
        \ExplainFuzz + JOIN & 2/3 & 27.4\%  & 2 & +7.44 & 2 & +7.44 \\
        \ExplainFuzz + GROUP & 0/4 & 11.8\%  & 0 & 0.00 & 0 & 0.00 \\
        \ExplainFuzz + UNION & 0/2 & 0.0\%  & 0 & 0.00 & 0 & 0.00 \\
        \midrule
        \textbf{Global result  } & \textbf{2/15} & \textbf{33.3\%} & \textbf{2} & \textbf{-3.46} &  \textbf{2} & \textbf{+76.37} \\
        \bottomrule
        \end{tabular}}
        \caption{Summary of conditioning evaluation for the SQL-1 domain.}
        \label{tab:sql1_summary}
    \end{subfigure}
    \caption{Effect of conditioning on \ExplainFuzz for the SQL-1 domain. 
    The heatmap (top) shows per-bug distinct triggers, while the table (bottom) summarizes alignment, coverage, and diversity improvements.}
    \label{fig:sql1_cond_overview}
\end{figure*}

\begin{figure*}[ht]
    \centering
    \begin{subfigure}[t]{0.95\textwidth}
        \centering
        \includegraphics[width=\textwidth]{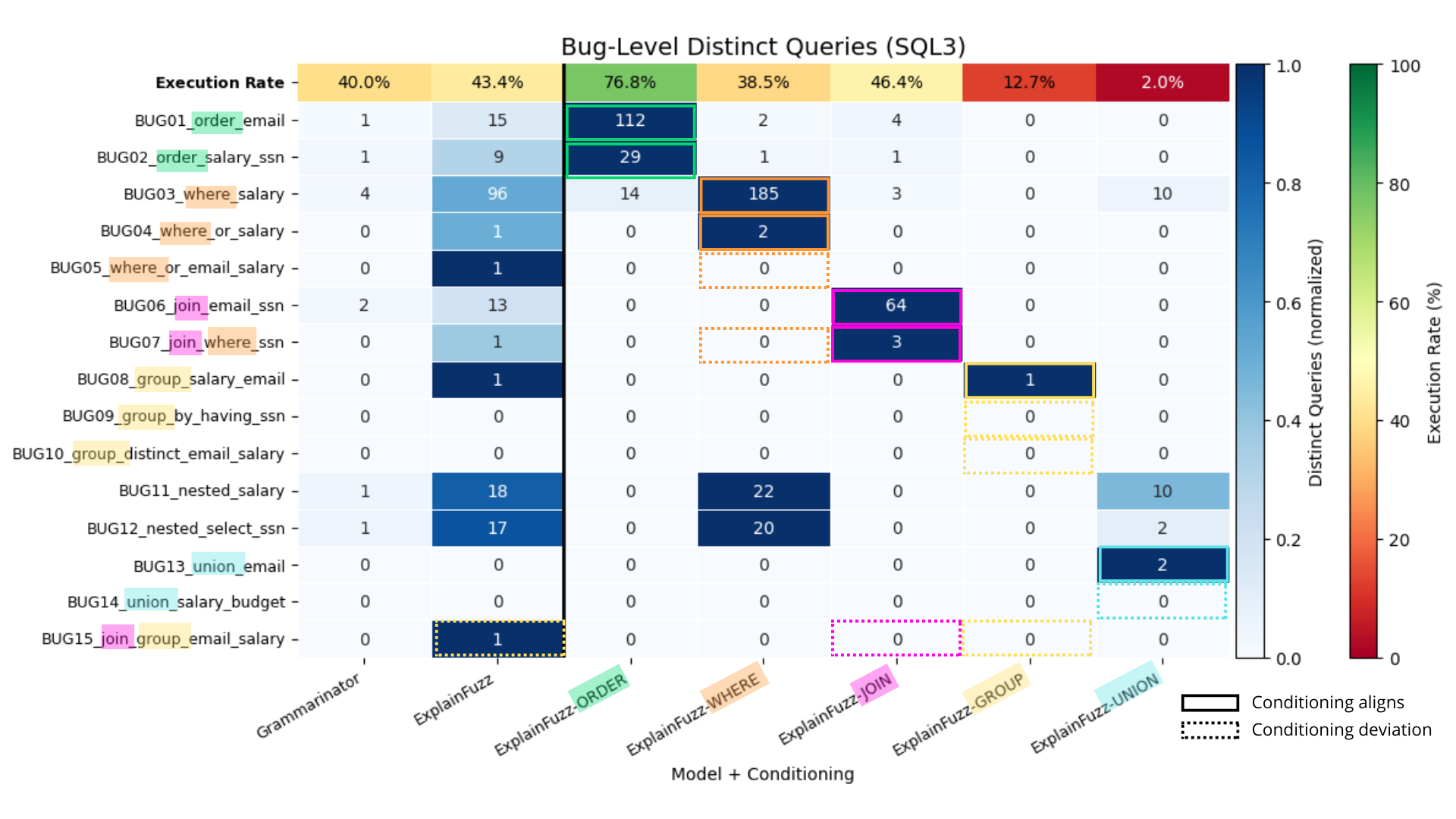}
        \caption{Bug-level heatmap showing the number of distinct queries triggering each bug for each model.}
        \label{fig:bug-level-heatm-SQL3}
    \end{subfigure}
    \vspace{0.5em}
    \begin{subfigure}[t]{0.95\textwidth}
        \centering
        \small
        \resizebox{\textwidth}{!}{
        \begin{tabular}{l >{\centering\arraybackslash}p{1.5cm} > {\centering\arraybackslash}p{1.6cm} >{\centering\arraybackslash}p{1.2cm} >{\centering\arraybackslash}p{1.8cm} >{\centering\arraybackslash}p{1.2cm} >{\centering\arraybackslash}p{1.8cm}}
        \toprule
        & & & \multicolumn{2}{c}{vs \ExplainFuzz} & \multicolumn{2}{c}{vs \Grammarinator} \\
        \cmidrule(lr){4-5} \cmidrule(lr){6-7}
        Conditioned Model & Alignment & Bug Coverage & New Bugs & Avg Diversity & New Bugs & Avg Diversity \\
        \midrule
        \ExplainFuzz + ORDER & 2/2 & 15.7\%  & 0 & +58.50 & 0 & +69.33 \\
        \ExplainFuzz + WHERE & 2/4 & 21.6\%  & 0 & +22.42 & 1 & +45.67 \\
        \ExplainFuzz + JOIN & 2/3 & 17.6\%  & 0 & +17.67 & 1 & +21.44 \\
        \ExplainFuzz + GROUP & 1/4 & 2.0\%  & 0 & 0.00 & 1 & +0.17 \\
        \ExplainFuzz + UNION & 1/2 & 7.8\%  & 1 & +0.67 & 1 & +0.67 \\
        \midrule
        \textbf{Global result  } & \textbf{8/15} & \textbf{66.7\%} & \textbf{1} & \textbf{+19.85} &  \textbf{4} & \textbf{+27.46} \\
        \bottomrule
        \end{tabular}}
        \caption{Summary of conditioning evaluation for the SQL-3 domain.}
        \label{tab:sql3_summary}
    \end{subfigure}
    \caption{Effect of conditioning on \ExplainFuzz for the SQL-3 domain. 
    The heatmap (top) shows per-bug distinct triggers, while the table (bottom) summarizes alignment, coverage, and diversity improvements.}
    \label{fig:sql3_cond_overview}
\end{figure*}

\begin{figure*}[ht]
    \centering
    \begin{subfigure}[t]{0.95\textwidth}
        \centering
        \includegraphics[width=\textwidth]{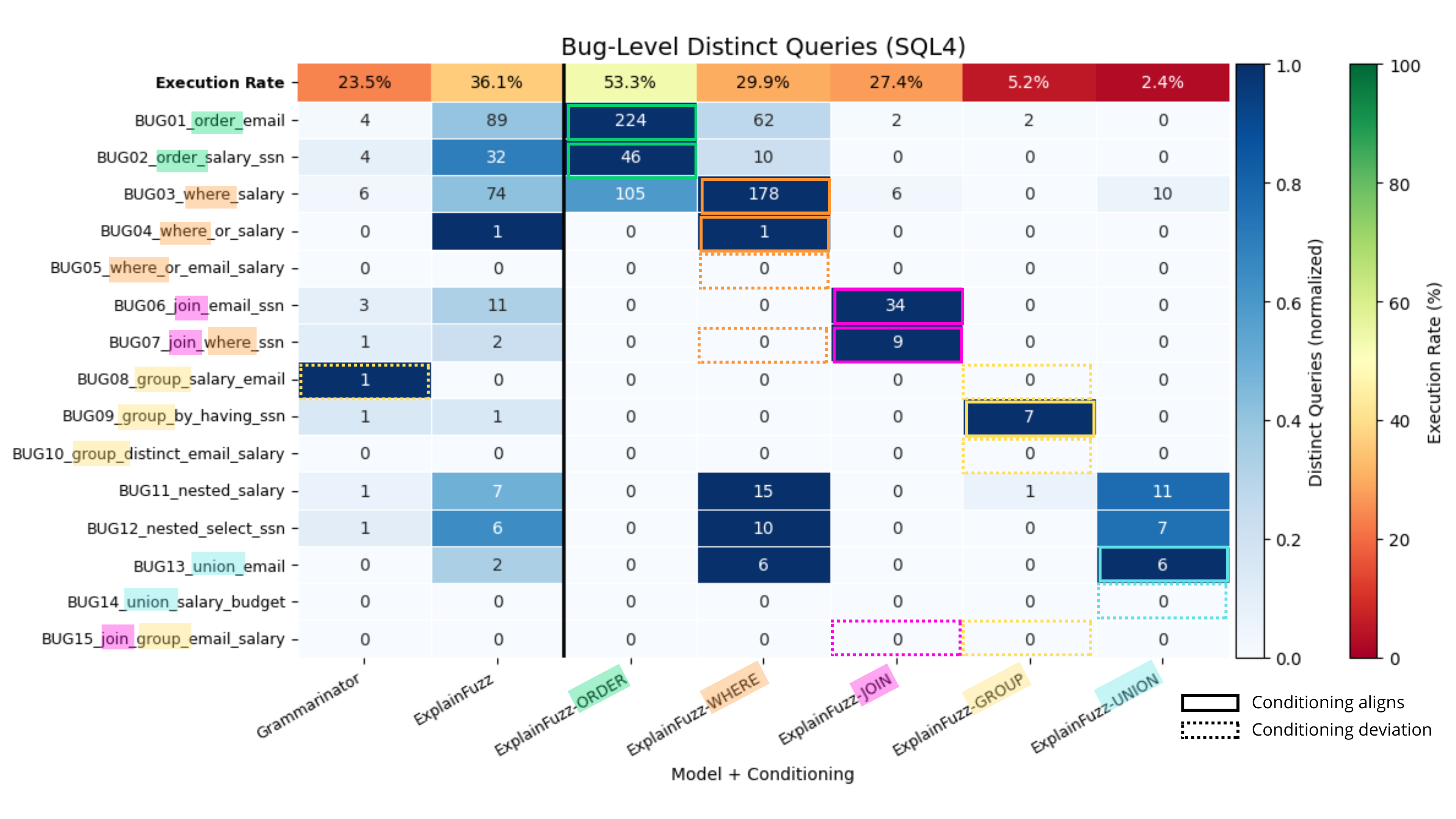}
        \caption{Bug-level heatmap showing the number of distinct queries triggering each bug for each model.}
        \label{fig:bug-level-heatm-SQL4}
    \end{subfigure}
    \vspace{0.5em}
    \begin{subfigure}[t]{0.95\textwidth}
        \centering
        \small
        \resizebox{\textwidth}{!}{
        \begin{tabular}{l >{\centering\arraybackslash}p{1.5cm} > {\centering\arraybackslash}p{1.6cm} >{\centering\arraybackslash}p{1.2cm} >{\centering\arraybackslash}p{1.8cm} >{\centering\arraybackslash}p{1.2cm} >{\centering\arraybackslash}p{1.8cm}}
        \toprule
        & & & \multicolumn{2}{c}{vs \ExplainFuzz} & \multicolumn{2}{c}{vs \Grammarinator} \\
        \cmidrule(lr){4-5} \cmidrule(lr){6-7}
        Conditioned Model & Alignment & Bug Coverage & New Bugs & Avg Diversity & New Bugs & Avg Diversity \\
        \midrule
        \ExplainFuzz + ORDER & 2/2 & 15.7\%  & 0 & +74.17 & 0 & +131.17 \\
        \ExplainFuzz + WHERE & 1/4 & 25.5\%  & 0 & +25.50 & 1 & +42.75 \\
        \ExplainFuzz + JOIN & 2/3 & 15.7\%  & 0 & +10.00 & 0 & +12.89 \\
        \ExplainFuzz + GROUP & 1/4 & 13.7\%  & 0 & +1.50 & 0 & +1.33 \\
        \ExplainFuzz + UNION & 1/2 & 23.5\%  & 0 & +2.17 & 1 & +3.00 \\
        \midrule
        \textbf{Global result  } & \textbf{7/15} & \textbf{66.7\%} & \textbf{0} & \textbf{+22.67} &  \textbf{2} & \textbf{+38.23} \\
        \bottomrule
        \end{tabular}}
        \caption{Summary of conditioning evaluation for the SQL-4 domain.}
        \label{tab:sql4_summary}
    \end{subfigure}
    \caption{Effect of conditioning on \ExplainFuzz for the SQL-4 domain. 
    The heatmap (top) shows per-bug distinct triggers, while the table (bottom) summarizes alignment, coverage, and diversity improvements.}
    \label{fig:sql4_cond_overview}
\end{figure*}

\begin{figure*}[ht]
    \centering
    \begin{subfigure}[t]{0.95\textwidth}
        \centering
        \includegraphics[width=\textwidth]{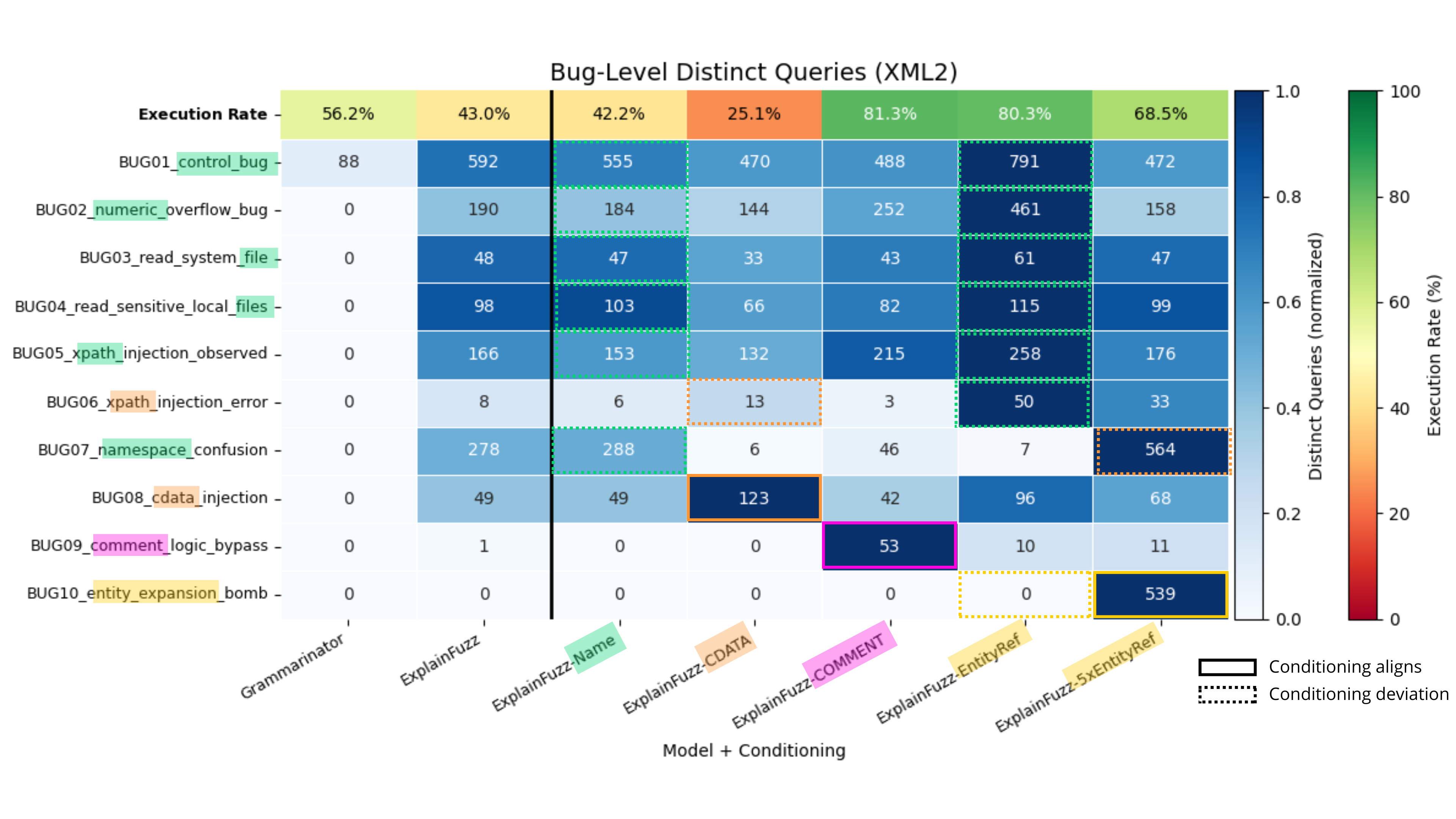}
        \caption{Bug-level heatmap showing the number of distinct queries triggering each bug for each model.}
        \label{fig:bug-level-heatm-XML2}
    \end{subfigure}
    \vspace{0.5em}
    \begin{subfigure}[t]{0.95\textwidth}
        \centering
        \small
        \resizebox{\textwidth}{!}{
        \begin{tabular}{l >{\centering\arraybackslash}p{1.5cm} > {\centering\arraybackslash}p{1.6cm} >{\centering\arraybackslash}p{1.2cm} >{\centering\arraybackslash}p{1.8cm} >{\centering\arraybackslash}p{1.2cm} >{\centering\arraybackslash}p{1.8cm}}
        \toprule
        & & & \multicolumn{2}{c}{vs \ExplainFuzz} & \multicolumn{2}{c}{vs \Grammarinator} \\
        \cmidrule(lr){4-5} \cmidrule(lr){6-7}
        Conditioned Model & Alignment & Bug Coverage & New Bugs & Avg Diversity & New Bugs & Avg Diversity \\
        \midrule
        \ExplainFuzz + Name & 2/6 & 80.0\%  & 0 & -7.06 & 5 & +206.67 \\
        \ExplainFuzz + COMMENT & 1/1 & 76.7\%  & 0 & +52.00 & 1 & +53.00 \\
        \ExplainFuzz + CDATA & 2/2 & 80.0\%  & 0 & +39.17 & 2 & +67.67 \\
        \ExplainFuzz + EntityRef & 0/1 & 86.7\%  & 0 & 0.00 & 0 & 0.00 \\
        \ExplainFuzz + 5xEntityRef & 1/1 & 96.7\%  & 1 & +538.33 & 1 & +538.33 \\
        \midrule
        \textbf{Global result  } & \textbf{6/11} & \textbf{100.0\%} & \textbf{1} & \textbf{+124.49} &  \textbf{9} & \textbf{+173.13} \\
        \bottomrule
        \end{tabular}}
        \caption{Summary of conditioning evaluation for the XML-2 domain.}
        \label{tab:xml2_summary}
    \end{subfigure}
    \caption{Effect of conditioning on \ExplainFuzz for the XML-2- domain. 
    The heatmap (top) shows per-bug distinct triggers, while the table (bottom) summarizes alignment, coverage, and diversity improvements.}
    \label{fig:xml2_cond_overview}
\end{figure*}

\begin{figure*}[ht]
    \centering
    \begin{subfigure}[t]{0.95\textwidth}
        \centering
        \includegraphics[width=\textwidth]{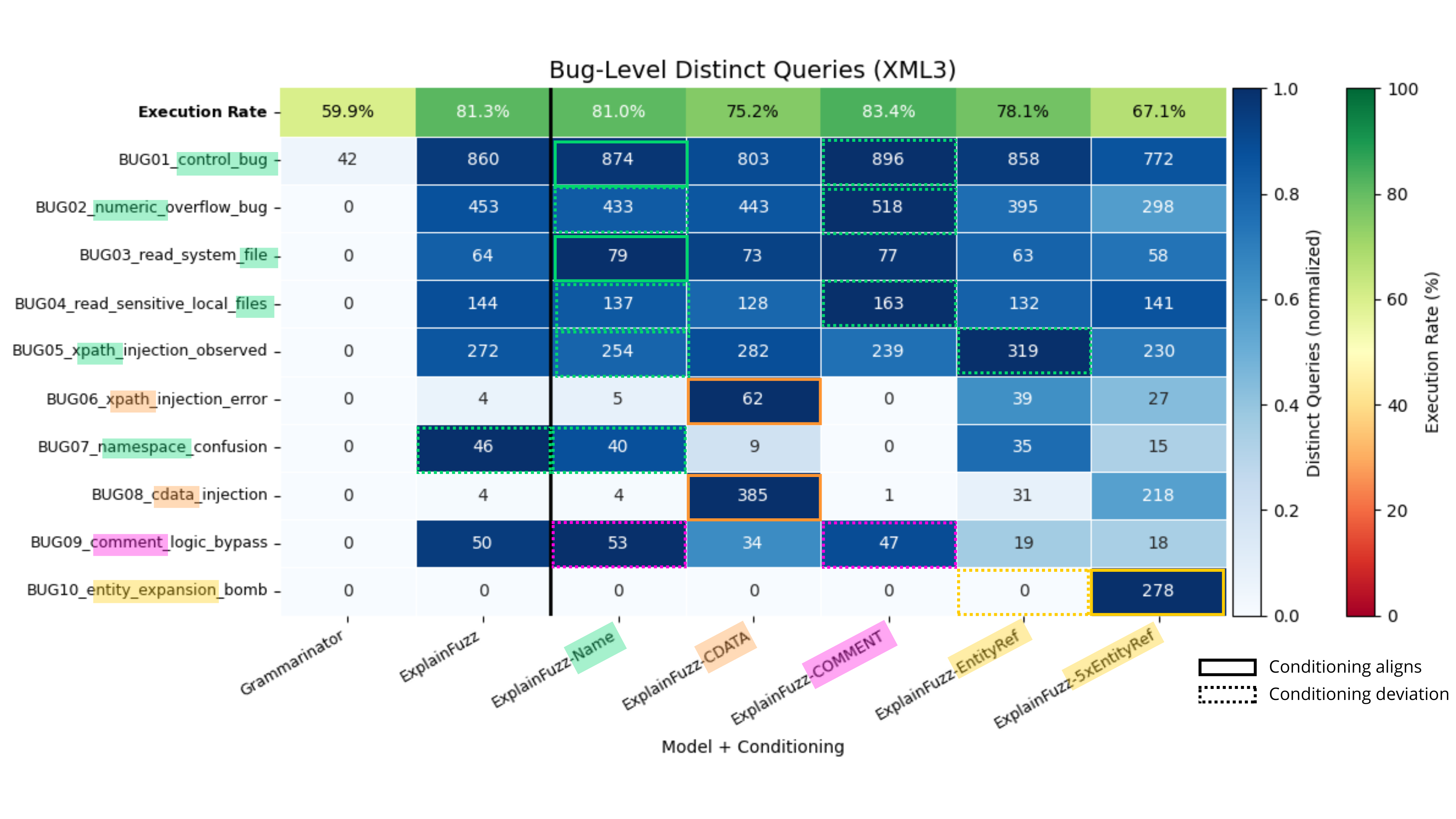}
        \caption{Bug-level heatmap showing the number of distinct queries triggering each bug for each model.}
        \label{fig:bug-level-heatm-XML3}
    \end{subfigure}
    \vspace{0.5em}
    \begin{subfigure}[t]{0.95\textwidth}
        \centering
        \small
        \resizebox{\textwidth}{!}{
        \begin{tabular}{l >{\centering\arraybackslash}p{1.5cm} > {\centering\arraybackslash}p{1.6cm} >{\centering\arraybackslash}p{1.2cm} >{\centering\arraybackslash}p{1.8cm} >{\centering\arraybackslash}p{1.2cm} >{\centering\arraybackslash}p{1.8cm}}
        \toprule
        & & & \multicolumn{2}{c}{vs \ExplainFuzz} & \multicolumn{2}{c}{vs \Grammarinator} \\
        \cmidrule(lr){4-5} \cmidrule(lr){6-7}
        Conditioned Model & Alignment & Bug Coverage & New Bugs & Avg Diversity & New Bugs & Avg Diversity \\
        \midrule
        \ExplainFuzz + Name & 2/6 & 90.0\%  & 0 & -3.61 & 5 & +295.50 \\
        \ExplainFuzz + COMMENT & 0/1 & 63.3\%  & 0 & -3.00 & 1 & +47.00 \\
        \ExplainFuzz + CDATA & 2/2 & 90.0\%  & 0 & +219.83 & 2 & +223.50 \\
        \ExplainFuzz + EntityRef & 0/1 & 90.0\%  & 0 & 0.00 & 0 & 0.00 \\
        \ExplainFuzz + 5xEntityRef & 1/1 & 100.0\%  & 1 & +277.33 & 1 & +277.33 \\
        \midrule
        \textbf{Global result  } & \textbf{5/11} & \textbf{100.0\%} & \textbf{1} & \textbf{+98.11} &  \textbf{9} & \textbf{+168.67} \\
        \bottomrule
        \end{tabular}}
        \caption{Summary of conditioning evaluation for the XML-3 domain.}
        \label{tab:xml3_summary}
    \end{subfigure}
    \caption{Effect of conditioning on \ExplainFuzz for the XML-3 domain. 
    The heatmap (top) shows per-bug distinct triggers, while the table (bottom) summarizes alignment, coverage, and diversity improvements.}
    \label{fig:xml3_cond_overview}
\end{figure*}

\begin{figure*}[ht]
    \centering
    \begin{subfigure}[t]{0.95\textwidth}
        \centering
        \includegraphics[width=\textwidth]{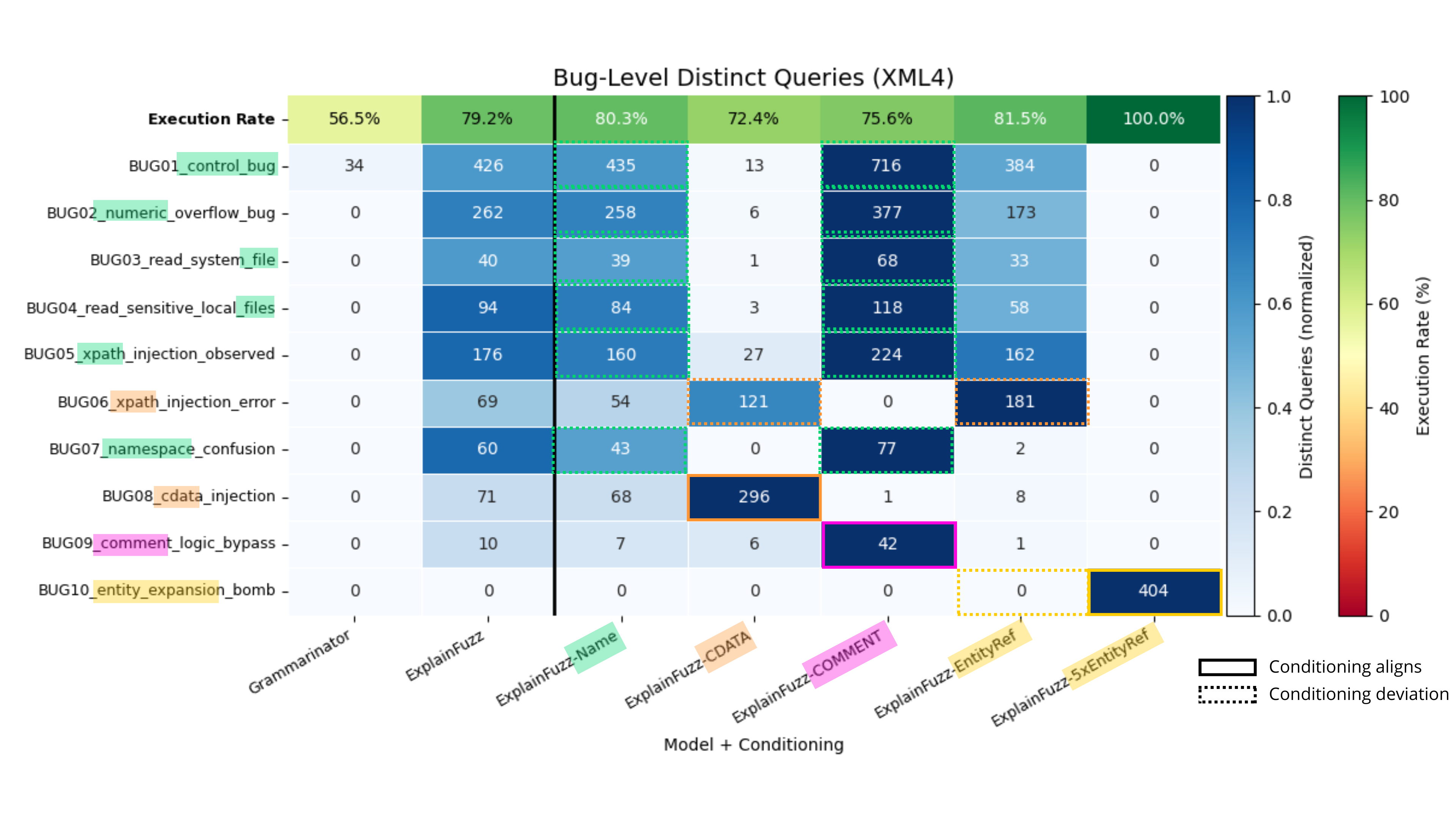}
        \caption{Bug-level heatmap showing the number of distinct queries triggering each bug for each model.}
        \label{fig:bug-level-heatm-XML4}
    \end{subfigure}
    \vspace{0.5em}
    \begin{subfigure}[t]{0.95\textwidth}
        \centering
        \small
        \resizebox{\textwidth}{!}{
        \begin{tabular}{l >{\centering\arraybackslash}p{1.5cm} > {\centering\arraybackslash}p{1.6cm} >{\centering\arraybackslash}p{1.2cm} >{\centering\arraybackslash}p{1.8cm} >{\centering\arraybackslash}p{1.2cm} >{\centering\arraybackslash}p{1.8cm}}
        \toprule
        & & & \multicolumn{2}{c}{vs \ExplainFuzz} & \multicolumn{2}{c}{vs \Grammarinator} \\
        \cmidrule(lr){4-5} \cmidrule(lr){6-7}
        Conditioned Model & Alignment & Bug Coverage & New Bugs & Avg Diversity & New Bugs & Avg Diversity \\
        \midrule
        \ExplainFuzz + Name & 1/6 & 90.0\%  & 0 & -6.56 & 5 & +163.67 \\
        \ExplainFuzz + COMMENT & 1/1 & 73.3\%  & 0 & +32.33 & 1 & +42.00 \\
        \ExplainFuzz + CDATA & 2/2 & 66.7\%  & 0 & +138.50 & 2 & +208.00 \\
        \ExplainFuzz + EntityRef & 0/1 & 70.0\%  & 0 & 0.00 & 0 & 0.00 \\
        \ExplainFuzz + 5xEntityRef & 1/1 & 10.0\%  & 1 & +404.00 & 1 & +404.00 \\
        \midrule
        \textbf{Global result  } & \textbf{5/11} & \textbf{100.0\%} & \textbf{1} & \textbf{+113.65} &  \textbf{9} & \textbf{+163.53} \\
        \bottomrule
        \end{tabular}}
        \caption{Summary of conditioning evaluation for the XML-4 domain.}
        \label{tab:xml4_summary}
    \end{subfigure}
    \caption{Effect of conditioning on \ExplainFuzz for the XML-4 domain. 
    The heatmap (top) shows per-bug distinct triggers, while the table (bottom) summarizes alignment, coverage, and diversity improvements.}
    \label{fig:xml4_cond_overview}
\end{figure*}